\documentclass[aos,preprint]{imsart}

\RequirePackage[OT1]{fontenc}
\RequirePackage{amsthm,amsmath}
\RequirePackage[numbers]{natbib}

\usepackage[toc,page]{appendix}

\usepackage{amsmath}
\usepackage{graphicx,psfrag,epsf}
\usepackage{enumerate}
\usepackage{natbib}
\usepackage{url} 

\usepackage{amssymb,latexsym,amsmath}
\usepackage{enumerate}
\usepackage{graphicx}
\usepackage{epstopdf}

\usepackage{enumitem}

\usepackage{mathrsfs}
\usepackage[table]{xcolor}
\usepackage{caption}
\usepackage{subcaption}
\usepackage{color}
\usepackage{multirow}
\newtheorem{Properties}{Properties}

\setattribute{journal}{name}{}

\startlocaldefs
\numberwithin{equation}{section}
\theoremstyle{plain}
\endlocaldefs

\begin{document}

\begin{frontmatter}
\title{Valid and Approximately Valid Confidence Intervals for Current Status Data}
\runtitle{Valid and Approx. C.I.s for Current Status Data}

\begin{aug}
\author{\fnms{Sungwook} \snm{Kim}\thanksref{m1,m2}\ead[label=e1]{s.kim@usciences.edu}},
\author{\fnms{Michael P.} \snm{Fay}\thanksref{m1}\ead[label=e2]{mfay@niaid.nih.gov}}
\and
\author{\fnms{Michael A.} \snm{Proschan}\thanksref{m1}
\ead[label=e3]{proscham@niaid.nih.gov}
}

\thankstext{t2}{Supported by National Institute of Allergy and Infectious Diseases}
\runauthor{S.Kim, M.P.Fay, and M.A.Proschan}

\affiliation{National Institute of Allergy and Infectious Diseases\thanksmark{m1} and University of the Sciences in Philadelphia\thanksmark{m2}}

\address{Address of the First  author\\
Department of Mathematics, Physics, and Statistics\\
University of the Sciences in Philadelphia\\
600 South 43rd St.\\
Philadelphia - PA 19104\\
\printead{e1}\\
}

\address{Address of the Second and Third authors\\
Biostatistics Research Branch/DCR/NIAID\\
5601 Fishers Lane,\\
Rockville, MD 20852\\
\printead{e2}\\
\phantom{E-mail:\ }\printead*{e3}
}
\end{aug}

\begin{abstract}
We introduce a new framework for creating point-wise confidence intervals for the distribution of event times for current status data. Existing methods are based on asymptotics. Our framework is based on binomial properties and motivates confidence intervals that are very simple to apply and are valid, i.e., guarantee nominal coverage.
Although these confidence intervals are necessarily conservative for small sample sizes, asymptotically their coverage rate approaches the nominal one. This binomial framework also motivates approximately valid confidence intervals, and simulations show that these approximate intervals generally have coverage rates closer to the nominal level with shorter length than existing intervals, including the likelihood ratio-based confidence interval. Unlike previous asymptotic methods that require different asymptotic distributions for continuous or grid-based assessment,  the binomial framework can be applied to either type of assessment distribution.
\end{abstract}



\end{frontmatter}

\section{INTRODUCTION}
\label{sec:intro}

This paper is concerned with finding pointwise confidence intervals on the event time distribution, $F$, for current status data.
In current status data, the event time  is not observed, but we only know one assessment time for each individual and  whether the event for that individual has occurred by that time or not.   This type of data appears  in many animal studies, cross-sectional studies and quantal bioassay studies.
 For example, consider a lung cancer study in mice. In order to determine if the cancer has developed, the mice must be sacrificed. So with each mouse, we only know if the cancer event has occurred by the time of sacrifice or not. Another example is a  cross-sectional study of women to determine the distribution for age at onset of menopause. At her age at the survey time, each woman will have either reached menopause or not. A third example concerns quantal bioassay studies where we can assume a monotonic dose response and we expose $n$ animals to respective doses $d_{1},d_{2},\ldots, d_{n}$, and see if they live or die at that dose. Here dose acts as the ``time'' variable.
 Throughout the paper we assume that the assessment time for each individual is independent of the event. So for example, this assumption would be violated in the first example if we partially based the time of mouse sacrifice on the apparent health of the mouse.

Many papers (see below) have developed  point-wise confidence intervals (CI) for $F$,   but as far as we are aware, no one has studied valid CIs, ones that guarantee nominal coverage.  In this article, we introduce new point-wise CIs (valid CIs and approximately valid CIs) for $F$ and study their asymptotic properties.
The valid confidence interval has coverage rates greater than or equal to the nominal rate,
and its coverage  rates
asymptotically approach  the nominal rate if certain conditions are satisfied.
The approximate confidence interval does not guarantee the nominal rate, but its coverage rate will generally be closer to the nominal rate.

The nonparametric maximum likelihood estimate (NPMLE) of $F$, say $\hat{F}_n$, is relatively straightforward to calculate.
\cite{groeneboom1992information} show this and also introduce the limiting distribution of $\hat{F}_n-F$ in the current status model, when $G$, the distribution of the observed assessment times, is continuous. When the assessments are independent of events, the limiting distribution at a fixed time point $t$ is
\begin{equation}
\label{G_theorem}
n^{1/3}\left\{\hat{F}_{n}(t)-F(t)\right\}\overset{d}{\to} \left[\frac{4f(t)F(t)\{1-F(t)\}}{g(t)}\right]^{1/3}\mathbb{Z}\equiv \mathscr{C}\mathbb{Z}
\end{equation}
where
$f\geq 0$ is the derivative of $F$; $g\geq 0$ is the derivative of  $G$; $\mathbb{Z}\equiv \text{argmin}(W(t)+t^{2})$ and $W$ is two-sided Brownian motion starting from 0.
If $\mathscr{C}$ were known then a 100(1-$\alpha$)\% Wald-based CI for $F(t)$ would be given by
\begin{equation}
\label{G_CI}
\left[\hat{F}_{n}(t)-n^{-1/3} \mathscr{C}\mathbb{Z}_{(1-\alpha/2)}, \hat{F}_{n}(t)+n^{-1/3} \mathscr{C}\mathbb{Z}_{(1-\alpha/2)}\right]
\end{equation}
where $\mathbb{Z}_{(1-\alpha/2)}$ is the $100(1-\alpha/2)$th quantile of the limiting random variable  $\mathbb{Z}$.  \cite{groeneboom2001computing} showed how to compute the quantiles of  $\mathbb{Z}$, but $\mathscr{C}$ contains the unknown parameters, $F$, $f$ and $g$.
The distribution $F$ can be estimated with the NPMLE, but $f$ and $g$ are more difficult to estimate. They are usually estimated using kernel methods
\citep[see][]{banerjee2005confidence,choi2013practicable}, although parametric methods have also been proposed \citep[see][]{banerjee2005confidence}.
\cite{choi2013practicable} show that this method can be improved by using transformations. Despite the improvement, the coverage can be very poor
at the ends of the distribution and with smaller sample sizes. For example, \cite{choi2013practicable} show situations with simulated coverage rates of the transformed Wald-based CIs of less than 80\% for nominal 95\% confidence intervals with sample sizes as large as $100$.

A generally better method is to use a likelihood ratio-based test (LRT) for $F(t)$.
\cite{banerjee2001likelihood} introduced the LRT for $F$ in  current status data, derived its limiting distribution under continuous $G$, and then developed the CIs by inverting a series of point null hypothesis tests. Like the Wald-based CI (expression~\ref{G_CI}), the LRT CI has a non-standard asymptotic distribution, except this
 distribution does not depend on unknown parameters. Thus, the 95\% confidence interval only requires the 95$th$ percentile of that distribution.
 The LRT method only needs the NPMLE and restricted NPMLEs of $F(t)$.
Unfortunately, just as with  the Wald-based CI, the LRT CI can have lower coverage rates than the nominal rate at the edges of $F$ when the sample size is small. \cite{banerjee2005confidence} show through simulations that the LRT CIs perform better than the untransformed Wald-based CIs.
Although \cite{choi2013practicable} did not calculate LRT CIs for their simulations, they show that the transformed Wald-based CIs can have performance close to the LRT CIs in the case studied in \cite{banerjee2005confidence}. Because of this we use the LRT CIs as a benchmark.

\cite{banerjee2005score} introduced three score statistics for testing the hypothesis that $H_{o}:F(t)=\theta_{o}$
assuming continuous failure and assessment distributions. They
showed that the asymptotic distribution for all three score statistics is the same, but is different from those of the Wald statistics and LRT statistics.
Simulations showed that the Wald tests were generally less powerful than the score tests and LRT statistics, and one version of the score test
may have more power than the LRT in some situations. Despite these promising simulation results, as far as we are aware, the full development of the confidence intervals from the score tests and other systematic exploration of the properties of those score-based confidence intervals have not been done.
We will not discuss score-based confidence intervals further.

\cite{tang2012likelihood} considered the case where the examination times lie on a grid and multiple subjects can share the same examination time.
They discovered some interesting asymptotics based on defining the distance between grid points as $\delta(n)=cn^{-\gamma}$, which changes with sample size $n$.  The asymptotic distribution of the NPMLE converges to one of two distributions depending on whether $\gamma<1/3$ or $\gamma>1/3$, and has different behavior at the boundary. Furthermore, they developed an adaptive inference for $F(t)$ which does not require  the information about $\gamma$.
However, this method is restricted to the specific case of equally spaced grid points, so will not be discussed further.

The nonparametric bootstrap approach on the NPMLE has similar coverage problems as the transformed Wald-based method at the edges of the distribution
\citep[see][Table 1]{choi2013practicable}. A sub-sampling approach to the problem has been explored,  but it can have very poor coverage in
certain situations  \cite[see][Table 3]{banerjee2005confidence}.

Finally, there has been some recent theoretical work in smoothing maximum likelihood estimation assuming continuous assessment.
\cite{groeneboom2010maximum} suggested two alternative estimators of $F$ for current status data: maximum smoothed likelihood estimation (MSLE) and smoothed maximum likelihood estimation (SMLE).
\cite{groeneboom2010maximum} derived the asymptotically mean squared error (MSE) optimal bandwidth,
but that bandwidth depends on unknown nuisance parameters.
\cite[][Section 9.5]{groeneboom2014nonparametric} showed how to construct a SMLE-based CI for $F$ in current status data.
They generated  bootstrap samples with replacement from the original sample, then computed the bootstrap $(1-\alpha)$ intervals.  However, in this method, it is difficult to estimate the actual bias term sufficiently accurately.  Without the actual asymptotic bias term, the coverage rate may be lower than the nominal rate. We explored using this method 
\citep[see supplemental article][Figure~S.2]{kim2018valid}, but it is difficult to automatically choose the bandwidth, and the
coverage was not good.
\cite{hjort2009extending} showed that with certain regularity conditions, an empirical likelihood-based method can be used on a smoothed
survival estimate for current status data.


We propose  a new framework for current status CIs based on   binomial properties,
and  introduce both valid and approximately valid CIs within that framework.
The valid CIs  can be applicable to  both discrete and continuous distributions of $G$ with no distributional assumptions on $F$.
 The valid CIs guarantee coverage, at the cost of larger length of CIs.
 In the continuous case, the valid $100(1-\alpha)\%$ CI for $F(t)$ amounts to
 using the $m$ assessment times just before and just after  $t$ (if they exist), counting the number of times the event occurs before each of those $m$ assessments,
 and using those counts out of $m$ from the valid lower or upper binomial confidence limits as the CI on $F(t)$.
We show that in the continuous case under some regularity conditions, those
valid CIs are asymptotically accurate if and only if $m \rightarrow \infty$ and $m/n^{2/3} \rightarrow 0$ as $n \rightarrow \infty$.
Additionally, we show that we can get close to minimal widths when $m = n^{2/3}$.
If $F$ can be assumed smooth, then several approximate CIs are proposed that require
 estimates of the nuisance parameters ($F(t)$, $f(t)$ and $g(t))$.
 The best of the approximate CIs has  generally better coverage with shorter length intervals than the likelihood ratio-based CIs.

The rest of this article is organized as follows. In Section \ref{section2}, we introduce a class of valid CIs, and
  show a member of this class with asymptotically minimum length of the CI. Because those asymptotically minimum length CIs depend on unknown nuisance parameters, we perform calculations showing that a simple approximation depending only on sample size is close to the asymptotically minimum length CI in a variety of settings.  In Section \ref{section3}, we introduce approximate CIs based on the binomial framework (ABF CIs), and show conditions to  asymptotically approach the nominal coverage. Since these ABF CIs may not be monotonic in $t$, we suggest  adjustments for monotonicity.
     In Section~\ref{sectionSimulation} we perform simulations of three different scenarios, comparing three different types of CIs: valid CI, ABF CIs, and the LRT CI.
     Additionally, we perform extensive and systematic simulations comparing the LRT CI and the mid-$P$ ABF CI.
      In Section~\ref{sectionApplication} we apply these methods to hepatitis A data from Bulgaria (\cite{keiding1991age}).
      The conclusions are in Section \ref{conclusion}, and all proofs are in the appendix.

 \section{VALID CONFIDENCE INTERVALS}
\label{section2}
\subsection{General Class of Intervals}
\label{sec_GeneralClassofIntervlas}
We first define a  class of valid confidence intervals, and later consider subsets within that class with additional desirable properties besides validity.

Suppose the $n$ event times are independent identically distributed (iid) from distribution $F$, the assessment times are iid from distribution $G$, and the assessments are independent of the event times.  We index the assessments so they are ordered, writing them as $C_{1}\leq C_{2} \leq \cdots \leq C_{n}$, and we
 let $T_{1}, \cdots, T_{n}$ be the associated unobserved event
 times. Let $D_{i}=1$ if $T_i \leq C_i$ and $0$ otherwise, and
we only observe $C_{i}$ and $D_{i}$. The problem is to find a confidence interval for $F(t)$ for fixed $t$.

Our strategy is to use the monotonicity of $F$ and the fact that given $C_i$, the $D_i$ are independent Bernoulli with parameter $F(C_i)$.
For $a<b$, let $N(a,b)$ be the number of assessment times in the interval $[a,b]$, and $Y(a,b)$ be the number of deaths occurring by those $N(a,b)$ assessment times:
\begin{equation*}
N(a,b)=\sum_{i:C_{i}\in[a,b]} 1 \ \text{and}\  Y(a,b)=\sum_{i:C_{i}\in[a,b]} D_{i}.
\label{DefinitionYN}
\end{equation*}
We will use the assessment times in the interval $[a,t]$ to find a lower confidence limit for $F(t)$.

We relate $Y(a,t)$ to $B$,  a binomial random variable with parameters $\{N(a,t),F(t)\}$.
Write the $100(1-\alpha)\%$  valid central confidence interval on $F(t)$ for $B$ given fixed $N=N(a,t)$ as
 $\left( L\{1-\alpha/2; B, N\}, U\{1-\alpha/2; B, N\} \right)$. This is the usual valid (often called ``exact'')
 binomial  confidence
interval developed in \cite{clopper1934use}, and is the union of two one-sided $1-\alpha/2$ intervals so that the CI is central, meaning it is two-sided and the error is bounded by $\alpha/2$ on each side.
Exploiting the connection between the binomial and beta distributions, we can express these limits as follows.
Let $Be(q; v,w)$ be the $q$th quantile from a beta distribution with non-negative shape parameters $v$ and $w$,
and set $Be\{q;0,w\}=\text{point mass at 0};$ $ Be\{q;v,0\}=\text{point mass at 1}$.  The lower and upper limits for one-sided $100q\%$ confidence intervals are given by
\begin{equation}
\begin{array}{l l}
L\{q; B, N\}&=Be\{1-q; B, N-B+1\};\\
U\{q; B, N\}&=Be\{q; B+1, N-B\}.
\end{array}
\label{defn:LandU}
\end{equation}
\normalsize

Although $Y(a,t)$ is not binomial, it relates to $B$ in the following manner.
Let ${\bf C} \equiv \left[ C_1,\ldots,C_n \right]$ be the ordered assessment vector.
It is intuitively clear that for fixed $a \leq t$,
\[
\text{Pr}[Y(a, t)\leq y\,|\,{\bf C}]\geq \text{Pr}[B\leq y\,|\,{\bf C}]~ \text{for all $y$}
\]
with probability $1$, since for $C_{i} \in [a,t]$, $F(a) \leq F(C_{i}) \leq F(t)$. This implies that
\begin{small}
\begin{equation}
\label{lowerfixa}
\begin{array}{l}
\text{Pr}[L\{q;Y(a,t), N(a,t)\}\leq F(t)\,|\,{\bf C}]\geq q \mbox{ a.s.\ \ and\ \ } \\
\text{Pr}[L\{q;Y(a,t), N(a,t)\}\leq F(t)]\geq q,
\end{array}
\end{equation}
\end{small}
\noindent
where $0\leq q \leq 1$.
The proof is given in Appendix \ref{Appendix_equation_3}.  Note that (\ref{lowerfixa}) shows that the coverage probability is at least $q$, whether we think conditionally given the assessment times, or unconditionally by averaging over those assessment times.  Conditional coverage is important if one focuses on $t$s for which there are multiple assessment times nearby, for example. In that case it would no longer suffice to have the right coverage averaged over the assessment time distribution.

Analogously, for fixed $b \geq t$, we use the $N(t,b)$ assessment times in $[t,b]$ and the $Y(t,b)$ deaths by those times to form an upper confidence limit for $F(t)$:
\begin{small}
\begin{equation}
\label{upperfixb}
\begin{array}{l}
\text{Pr}[F(t)\leq U\{q;Y(t,b), N(t,b)\}\,|\,{\bf C}]\geq q \mbox{ a.s.\ \ and\ \ } \\
\text{Pr}[F(t)\leq U\{q;Y(t,b), N(t,b)\}]\geq q.
\end{array}
\end{equation}
\end{small}
\noindent

\begin{figure}[t]
\centering
\includegraphics[height=3in,width=4in]{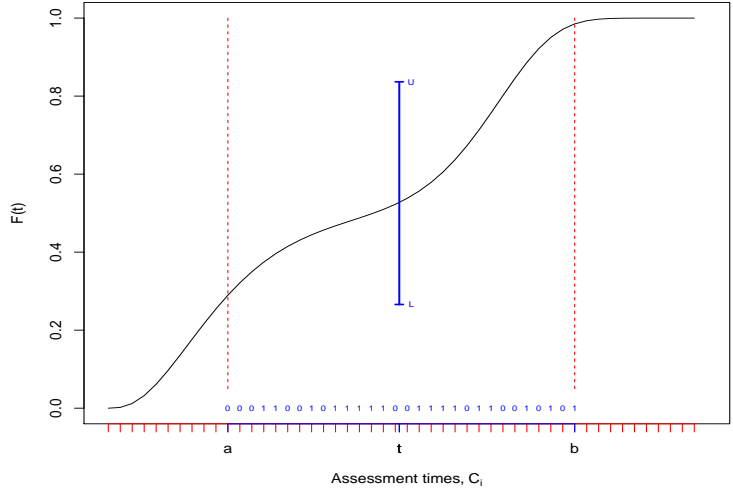}
 \caption{\textit{An example of the valid two sided confidence interval about $F(t)$, (\ref{CPtwoside}). $Y(a,t)=8$, $N(a,t)=15$, $Y(t,b)=9$, and $N(t,b)=15$.}}
  \label{ExampleCI}
\end{figure}

Then for $a \leq t \leq b$, a valid central 100(1-$\alpha$)\%   confidence interval about $F(t)$ can be formed by combining the one-sided limits from inequalities (\ref{lowerfixa}) and (\ref{upperfixb}):
\begin{eqnarray}
\left[ L \{1 -\alpha/2; Y(a,t), N(a,t) \}, \right.
& &
\left.
U \{1-(\alpha/2); Y(t, b), N(t,b) \} \right]  \label{CPtwoside}
\end{eqnarray}
We plot an example of one of these intervals in Figure \ref{ExampleCI}.

We mentioned earlier that one might focus on certain time points after observing the $C$s.  In other words, the $a$ and $b$ might not be constants fixed in advance, but functions of the $C$s.  The following theorem shows that this does not cause a problem.

\textsc{Theorem} 1a. \textit{Let $a$ and $b$ be known functions of only $t$, $n$, and $\bold{C}=(C_1,\ldots,C_n)$,  such that  $a(t,n,\bold{C})\leq t \leq b(t,n,\bold{C})$ with probability $1$. Let $Y_{a}^t = Y(a(t,n,\bold{C}), t)$, $Y_{t}^b = Y(t,b(t,n,\bold{C}))$, and similarly for $N^{t}_{a}$ and  $N^{b}_{t}$. If
$L=L \left\{ 1-\alpha/2;Y^{t}_{a},N^{t}_{a} \right\}$ and $U=U \left\{1-\alpha/2;Y^{b}_{t}, N^{b}_{t} \right\}$, then
\begin{eqnarray}
& & \text{Pr} \left[L\leq F(t)\leq U \,|\,{\bf C} \right] \geq 1-\alpha {\rm\ a.s.\ and\ \ } \text{Pr} \left[L \leq F(t)\leq U\right] \geq 1-\alpha, \label{twoside}
\end{eqnarray}
for any $n$, and additionally $(L,U)$ is central.}\\

The proof is given in Appendix \ref{Appendix_Proof_1a}.

Before discussing specific forms of the functions $a$ and $b$, note that it is possible that $L>U$, where $L$ and $U$ are the lower and upper limits given in Theorem 1a. When $L>U$, we have the freedom to redefine the limits to whatever we want without violating validity.  The redefined limits can even depend on $Y_a^t$ and $Y_t^b$.\\

\textsc{Theorem} 1b. \textit{
Let $L \equiv L \left\{ 1-\alpha/2;Y^{t}_{a},N^{t}_{a} \right\}$ and
$U \equiv U \left\{1-\alpha/2;Y^{b}_{t}, N^{b}_{t} \right\}$ as defined in Theorem 1a, with $a(t,n,\bold{C})\leq t \leq b(t,n,\bold{C})$.
Let
\begin{equation*}
\label{LUstar}
\begin{array}{l}
L^{*}= \left\{
  \begin{array}{l l}
L & \quad \text{if}~L\leq U\\
L_{M}  & \quad \text{if}~ L>U;
  \end{array} \right.\\
$~$\\
U^{*}= \left\{
  \begin{array}{l l}
U & \quad \text{if}~L\leq U\\
   U_{M} & \quad \text{if}~ L>U.
  \end{array} \right.
\end{array}
\end{equation*}
where $L_M \leq U_M$ can be any statistics.
Then
\begin{equation}
\label{twoside_with_m}
\text{Pr}\{ L^{*}\leq F(t)\leq U^{*}\} \geq 1-\alpha
\end{equation}
for any $n$.}\\

The result follows immediately from the fact that whenever the original interval $[L,U]$ covers $F$, so does the modified interval.

Although setting $L_{M}= U_{M}=\hat{F}(t)$ will give the minimum length CI for  $L_{M}$ and $U_{M}$, given  $a(t,n,\bold{C})$ and $b(t,n,\bold{C})$, these intervals are not practical since in the continuous case, $\text{Pr}\left\{F(t) \in \left[L_{M}, U_{M}\right]| L_{M}=U_{M}\right\}=0$, and most users of the confidence interval would not accept $L_{M}=U_{M}$ when $L>U$ because of that zero conditional coverage.

In the next section we discuss choosing from among those intervals of Theorem 1a or 1b, some of which can have very wide expected length.

\subsection{Asymptotic Properties  for  Nominal Coverage}
\label{sec_APNC}

In this section, we first introduce a specific form of the functions $a$ and $b$, and then discuss conditions so that the asymptotic coverage goes to the nominal level.

For a fixed $m$ given $n$,  we consider two random  points $a$ and $b$ defined by the $C_i$. Starting at point $t$, go backward in time to the $m$th closest $C_i$ less than or equal to $t$ (or backward to $0$ if there are fewer than $m$ points less than or equal to $t$).  Denote that point as $a=a(t,n,\bold{C})$.  Similarly, go forward in time from $t$ to find the $m$th closest $C_i$ greater than or equal to $t$ (or $\infty$ if fewer than $m$ points are greater than or equal to $t$).  More formally,
\begin{equation}
\label{definitionab}
\begin{array}{l}
a\equiv a(t,n,\bold{C})= \left\{
  \begin{array}{l l}
    0& \quad \text{if $C_{m}>t$ }\\
    C_{l-m+1} & \quad \text{if $C_{m}\leq t$};
  \end{array} \right.\\
$~$\\
b\equiv b(t,n,\bold{C})= \left\{
  \begin{array}{l l}
    \infty& \quad \text{if $C_{n-m+1}<t$ }\\
    C_{g+m-1} & \quad \text{if $C_{n-m+1}\geq t$,}
  \end{array} \right.
\end{array}
\end{equation}
where $m=m(n)$ is a function of $n$ only, and $m$ is a positive integer,  $l=\text{max}\{i:C_{i}\leq t\}$, and $g=\text{min}\{i:C_{i}\geq t\}$. For convenience, let $C_{0}=0$ and $C_{n+1}=\infty$. If there are ties at $a(t,n,\bold{C})$ or $b(t,n,\bold{C})$ then we include all ties. Therefore there are at least  $m$  observations within $[a(t,n,\bold{C}),t]$ and  within $[t,b(t,n,\bold{C})]$ when $C_{m}\leq t \leq C_{n-m+1}$. If $G$ is continuous and $a(t,n,\bold{C})\neq 0$, then $N\{a(t,n,\bold{C}),t)\}=m$ with probability $1$. Analogously, if $G$ is continuous and $b(t,n,\bold{C})\neq \infty$, then $N\{b(t,n,\bold{C}),t)\}=m$ with probability $1$.

As was described in Section \ref{sec_GeneralClassofIntervlas},  it is possible that $L>U$.
If $L>U$, then we use $L_{M}\equiv L\{1-(\alpha/2);Y_{a^{*}}^{ b^{*}}, N_{a^{*}}^{b^{*}}\}$ and $U_{M}\equiv U\{1-(\alpha/2);Y_{a^{*}}^{ b^{*}}, N_{a^{*}}^{b^{*}}\}$, where $a^{*}\neq b^{*}$ are specified in
\cite[][Section~S.1]{kim2018valid}.
Essentially, instead of using separate proportions of $m$ observations less than $t$ and $m$ observations greater than $t$ to form the lower and upper confidence limits, we use a single proportion combining $m/2$ observations less than, and $m/2$ observations greater than, $t$.

With these specific forms of the functions $a$ and $b$, the confidence interval  constructed with any $m$  is valid.
 An additional desirable property on the function $m(n)$ is that the resulting confidence intervals are asymptotically accurate, meaning that the the coverage probability converges to the desired level. This property can be met
 at the support of $G$ for discrete $G$ if $m(n) \equiv m_n$ has the following two conditions:

\begin{Properties}
$~$\\
\item $\lim_{n\rightarrow \infty}(m_n/n)=0$;\\
\item  $\lim_{n \rightarrow \infty}(m_n)=\infty$.
\end{Properties}


\textsc{Theorem 2.1.}  \textit{If
Properties~1
are satisfied,  and  $G$ is discrete, then
the coverage rate of both (\ref{twoside}) and  (\ref{twoside_with_m}) are $1-\alpha$ as $n\rightarrow \infty$ at each atom of $G$.}\\

 The proof is given in Appendix \ref{Appendix_Proof_2a}.

Hereafter, we assume that $G$ is continuous. If $C_{m_{n}}\leq t$ then
\begin{equation}
\label{intB}
N_a^t = m_n \mbox{ and }
Y^{t}_{a}=\sum^{l_{n}}_{i=l_{n}-m_{n}+1} D_{i},
\end{equation}
and if $a(t,n,\bold{C})\leq C_{i}\leq t$ then $D_{i}|C_{i}\sim \text{Bernoulli}\left\{ F(C_{i})\right\}$ where $F\{a(t,n,\bold{C})\}\leq F(C_{i})\leq F(t)$ for $i=(l_{n}-m_{n}+1)\ldots l_{n}$.  We have noted previously that the conditional distribution of $Y_a^t$ given ${\bf C}$ is stochastically between a binomial $(m_n,F(a_n))$ and a binomial $(m_n,F(t))$. If $a_n\rightarrow t$ fast enough, we should be able to approximate both of these binomial distributions with normals with means $m_nF(t)$ and variances $m_nF(t)\{1-F(t)\}$, which would guarantee asymptotic accuracy of the lower confidence limit, and similarly for the upper limit.  We seek conditions under which this holds.

Let $W_{m_n}$ and $W_{m_n}^\prime$ denote binomials with parameters $(m_n,F(t))$ and $(m_n,F(a_n))$, respectively.  By the central limit theorem, $Z_n=\{W_{m_n}-m_nF(t)\}/[m_nF(t)\{1-F(t)\}]^{1/2}$ converges in distribution to a standard normal.  Call $Z_n^\prime=\{W_{m_n}^\prime-m_nF(t)\}/[m_nF(t)\{1-F(t)\}]^{1/2}$ the {\it lower standardized deviate}.  Similarly, if $W_{m_n}^{\prime\prime}$ denotes a binomial with parameters $(m_n,F(b_n))$, call $Z_n^{\prime\prime}=\{W_{m_n}^{\prime\prime}-m_nF(t)\}/[m_nF(t)\{1-F(t)\}]^{1/2}$ the {\it upper standardized deviate}.  We want to know when the lower and upper standardized deviates are both asymptotically standard normal, which would gurantee asymptotic accuracy.\\

\textsc{Theorem 2.2.}  \textit{Assume that $F$ and $G$ are continuous and, at the point $t$, $F^\prime(t)=f(t)>0$ and $G^\prime(t)=g(t)>0$.  Assume further that
$m_n\rightarrow \infty$ and $m_n/n\rightarrow 0$.  Then the lower and upper standardized deviates converge in distribution to standard normals (which guarantees that the conditional and unconditional coverage both tend to $1-\alpha$ as $n\rightarrow \infty$) if and only if $m_n/n^{2/3}\rightarrow 0$ as $n\rightarrow \infty$.}\\

 The proof is given in Appendix \ref{Appendix_Proof_2b_i}.

\mathversion{bold}
\subsection{Choice of $m_{n}$}
\label{Sec_Opti_Min_Length}
\mathversion{normal}

Although Theorems 2.1 and 2.2 give conditions on $m_n$ that lead to asymptotically accurate coverage, there is quite a range of functions $m(n)$ that lead to
 asymptotic accuracy. Further, since Theorem~1 shows guaranteed nominal coverage
for a wider class of intervals, within this wider class the only error in coverage will be higher (i.e., better) coverage.
So practically speaking, for choosing $m_n$ we focus in this section not on coverage, but on minimum expected length.

We motivate a simple $m(n)$ function using three steps. First, we motivate more accurate binomial approximations for $Y_{a}^{t}$ and $Y_{t}^{b}$
than those used in Theorem 2.2. These approximations are based on $n$, $F(t)$ and $r(t) = f(t)/g(t)$ only.
Second, through numerical search we find the $m(n)$ that gives the lowest expected length 95\% confidence interval for several $n$, $F(t)$ and $r(t)$ values.
Third, we show that $m(n) = n^{2/3}$ is close to that minimum when $r(t)=1$, and the expected length is close to the minimum expected length for $1/2 < r(t) < 2$.

In Theorem 2.2, we approximated the distribution of $Y_a^t$ by a binomial with parameters $(m_n,F(t))$.  The following heuristic argument gives a more accurate approximation to the distribution function for $Y_a^t$.

Assuming $G$ is continuous, for  $\bold{C}$ fixed and $C^* \sim G$ for all $j$, $m_{n}$ is approximately $n$ $\times$ Pr\{$a(t,n,\bold{C})\leq C^* \leq t$\}=$n[G(t)-G\{a(t,n,\bold{C})\}]$.
Also, $G\{a(t,n,\bold{C})\}=G(t)-\{t-a(t,n,\bold{C})\}g(t)+o\{t-a(t,n,\bold{C})\}$ as $a(t,n,\bold{C})\rightarrow t$ because $G^\prime(t)=g(t)$.  Using the approximation $G(t)-G\{a(t,n,\bold{C})\}\approx g(t)\{t-a(t,n,\bold{C})\}$, we can write $m_{n}$ as $m_{n}\approx ng(t)\{t-a(t,n,\bold{C})\}$, implying that
\begin{equation}
\label{eps1}
\{t-a(t,n,\bold{C})\}\approx \frac{m_{n}}{ng(t)}.
\end{equation}
Likewise, $F\{a(t,n,\bold{C})\}=F(t)-\{t-a(t,n,\bold{C})\}f(t)+o\{a(t,n,\bold{C})\}$ as $a(t,n,\bold{C})\rightarrow t$, so
\begin{equation}
\label{Feps}
F\{a(t,n,\bold{C})\}\approx F(t)-\{t-a(t,n,\bold{C})\} f(t)
\end{equation}
for large $n$.
Using (\ref{eps1}) and (\ref{Feps}), we can approximate $F\{a(t,n,\bold{C})\}$ for large $n$ as follows:
\begin{equation}
\label{Bin1}
F\{a(t,n,\bold{C})\}\approx  F(t)-\left \{\frac{m_{n}}{ng(t)}\right \} f(t).
\end{equation}
Analogously, with approximations similar to those used for (\ref{Bin1}), $F\{b(t,n,\bold{C})\}$ can be written as
\begin{equation}
\label{Bin2}
F\{b(t,n,\bold{C})\}\approx F(t)+\left \{\frac{m_{n}}{ng(t)}\right \} f(t).
\end{equation}

Then we approximate the distribution of $Y_{a}^{t}$ as
\begin{equation}
\label{approxYat}
Y_{a}^{t}~\dot{\sim}~ \text{Binomial}(m_{n}, F^{-}_{t})
\end{equation}
where $F^{-}_{t}$ is the midpoint of $F\{a(t,n,\bold{C})\}$ and $F(t)$:
\begin{equation}
\label{Fminus}
F^{-}_{t}=\frac{F\{a(t,n,\bold{C})\}+ F(t)}{2}.
\end{equation}
Using (\ref{Bin1}) and (\ref{Fminus}), we can express (\ref{approxYat}) as
\begin{equation}
\label{approxYat2}
Y_{a}^{t}~\dot{\sim}~ \text{Binomial}\left[m_{n},F(t)-\left\{\frac{m_{n}}{2ng(t)}\right\} f(t)\right].
\end{equation}
Analogously,
\begin{equation}
\label{approxYtb2}
Y_{t}^{b}~\dot{\sim}~ \text{Binomial}\left[m_{n},F(t)+\left\{\frac{m_{n}}{2ng(t)}\right\} f(t)\right].
\end{equation}

In Table~\ref{TabMinM} we give $m_{min}$, the $m_n$ that gives the minimum expected 95\% confidence interval length using the Clopper-Pearson intervals
associated with approximations
 (\ref{approxYat2}) and (\ref{approxYtb2}) for different values of $F(t)$, $r(t)$ and $n$.
Given $m_n$ we calculate the expected 95\% confidence interval length by subtracting the weighted average of the $m_n+1$ possible values for $U(0.975; Y_t^b,m_n)$ from those for $L(0.975; Y_a^t, m_n)$, weighted by the appropriate binomial probabilities (see expressions~\ref{approxYat2} and \ref{approxYtb2}).
We find  $m_{min}$  by exhaustive computer search.
We see that for $r(t)=1$ it appears that $\lceil n^{2/3} \rceil$ is a good estimator of $m_{min}$. For $r(t) \neq 1$ then $\lceil n^{2/3} \rceil$ is a much poorer estimator.
However, even though $\lceil n^{2/3} \rceil$ is not close to $m_{min}$, we find that the expected 95\% confidence interval length is not too much inflated by using the suboptimal
$\lceil n^{2/3} \rceil$ for $m_n$.
Table~\ref{TabMinM} gives  the ratio of the expected 95\% confidence interval length when $m=\lceil n^{2/3} \rceil$ over the expected length
at $m_n = m_{min}$, and we see that the expected inflation is 8\% or less for all the situations explored ($r(t)=1/2, r(t)=1$ and $r(t)=2$).
The same calculations for 90\% confidence intervals have similar $m_{min}$ and  expected inflation of 12\% of less for the same explored situations (not shown).

\begin{table}[ht]
\centering
\caption{
For different values of $n$ (first column) we give $n^{2/3}$ rounded up to the nearest integer (second column). The other columns
give  $m_{min} (E_{ratio})$, where $m_{min}$ is the value of $m_n$  that gives estimated minimum expected 95\% confidence interval length,
and $E_{ratio}$ is the ratio of expected 95\% CI length when $m_n = \lceil n^{2/3} \rceil$ over the expected CI length when $m_n = m_{min}$.
Estimations are based on exhaustive calculations assuming the binomial approximations (\ref{approxYat2}) and (\ref{approxYtb2})
are exactly correct.
\label{TabMinM}
}
\begin{tabular}{rrrrrrrr}
  \hline
  & & \multicolumn{2}{c}{r(t)=1}  & \multicolumn{2}{c}{r(t)=.5} & \multicolumn{2}{c}{r(t)=2} \\
n & $\lceil n^{2/3} \rceil$ & F(t)=0.5 & F(t)=0.75 & F(t)=0.5 & F(t)=0.75 & F(t)=0.5 & F(t)=0.75 \\
  \hline
100 & 22 & 22 (1.00) & 21 (1.00) & 31 (1.04) & 31 (1.03) & 13 (1.06) & 13 (1.05) \\
  200 & 35 & 35 (1.00)  & 33  (1.00) & 53 (1.05) & 52 (1.03) & 22 (1.06) & 21 (1.06) \\
  500 & 63 & 65 (1.00) & 60 (1.00) & 103 (1.06) & 95 (1.04) & 41 (1.05) & 38 (1.07) \\
  1000 & 100 & 103 (1.00) & 95 (1.00) & 162  (1.06) & 149 (1.04) & 65 (1.05) & 60 (1.07) \\
  2000 & 159 & 162  (1.00) & 149 (1.00) & 257 (1.05) & 235 (1.04) & 103  (1.05) & 95  (1.07) \\
  5000 & 293 & 297 (1.00) & 272 (1.00) & 470  (1.05) & 430 (1.04) & 188  (1.05) & 173 (1.08) \\
  10000 & 465 & 470  (1.00) & 430 (1.00) & 743 (1.05) & 678 (1.03) & 297  (1.05) & 272  (1.08) \\
   \hline
\end{tabular}
\end{table}

\begin{figure}[t!]
    \centering
    \begin{subfigure}[t]{0.5\textwidth}
        \centering
        \includegraphics[height=2.5in,width=2.5in]{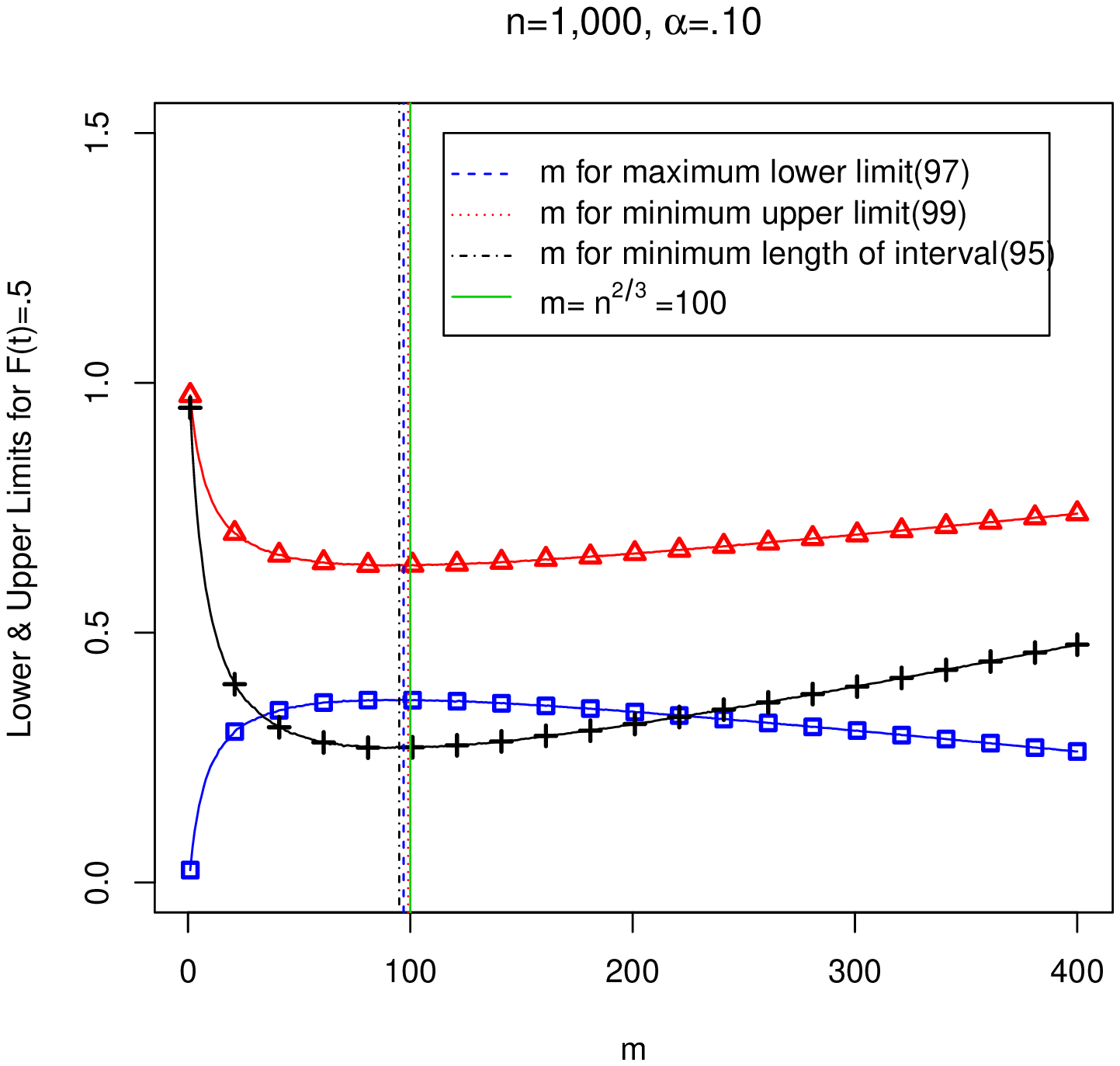}
        \caption{90\% confidence interval with $m$s}
\label{ac90m}
    \end{subfigure}%
    ~
    \begin{subfigure}[t]{0.5\textwidth}
        \centering
        \includegraphics[height=2.5in,width=2.5in]{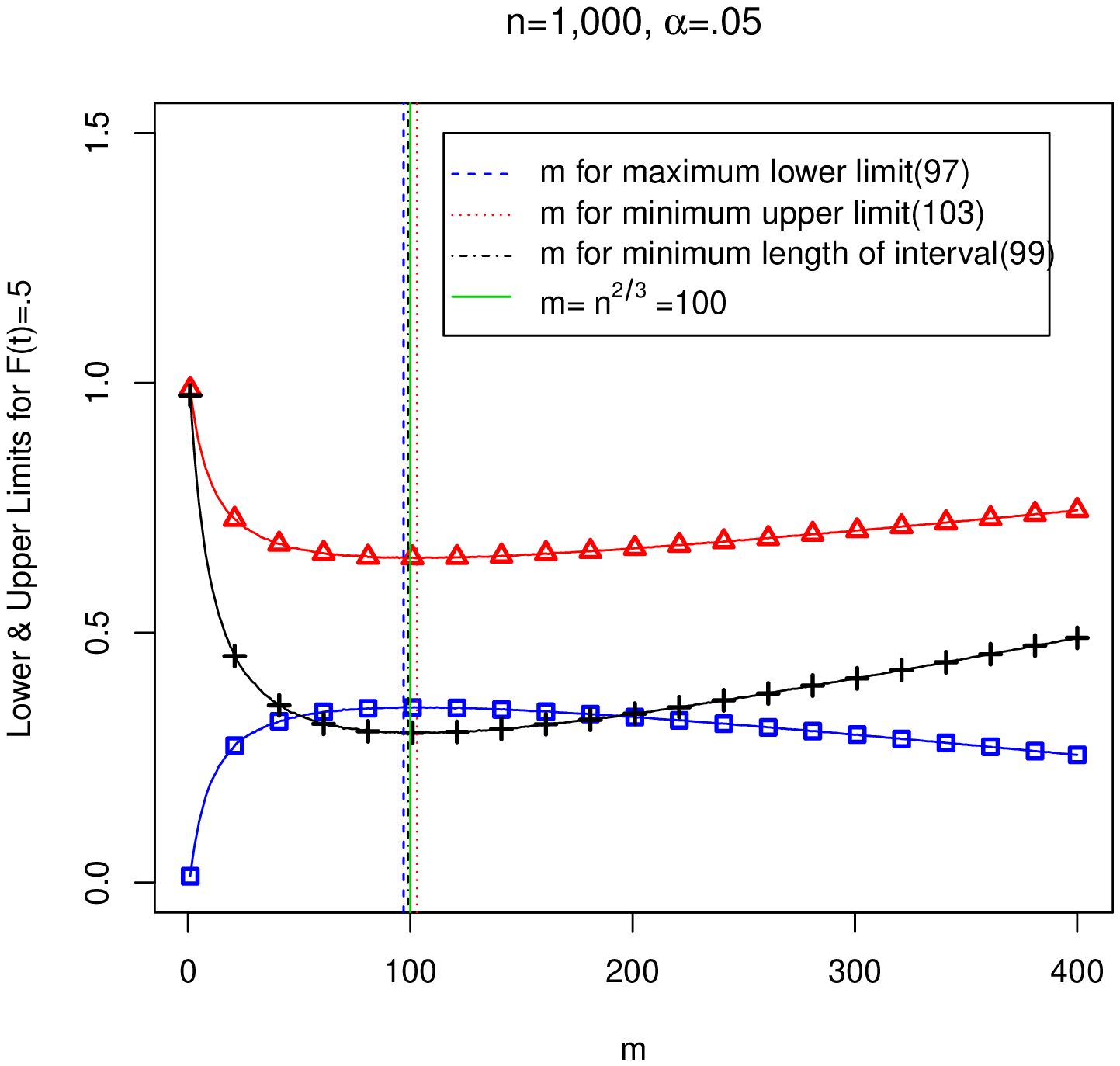}
        \caption{95\% confidence interval with $m$s}
\label{ac95m}
    \end{subfigure}
    \caption{\textit{The average of 10,000 simulated confidence intervals about $F(t)$=.5 with various $m$s. The average of lower
    limits: blue solid line with ($\Box$); the average of the length of confidence intervals: black solid line with ($+$);
    the average of upper limits: red solid line with ($\triangle$).}}
\label{Mlengthfigure}
\end{figure}

To explore how well the approximation does in picking $m_{min}$,
we simulated 10,000 confidence intervals at $F(t)=.5$ when $F=G$
with $n=1,000$. For the simulations we used $F=G$ are both exponential with mean 1, but
because the relationship between $F$ and $G$ is all that matters in the continuous case, we would get the same results
when both $F$ and $G$ are the same continuous distribution.
Figure \ref{Mlengthfigure} shows the average length of the CIs of 10,000 simulated confidence intervals
with various $m$s ($m=1,\ldots, 400$). We see that the value $m_n$ that gives minimum simulated confidence interval length ($m_{n}=95$ for 90\% confidence level, and $m_{n}=99$ for 95\% confidence level)   is close to $n^{2/3}=100$, and that the expected
length does not change much around that value.

\section{CONFIDENCE INTERVALS WITH COVERAGE CLOSE TO NOMINAL}
\label{section3}
\subsection{Notation and a Theorem}

Up to now, we have considered a conservative valid method which, for continuous assessments away from the boundaries (see (\ref{definitionab}) and discussion afterward), uses the $m_{n}$ observations closest to $t$ and less than or equal to $t$
 for the lower limit and analogously uses $m_{n}$ observations closest to $t$ and greater than or equal to $t$ for the upper limit.  In this section, we construct less conservative confidence intervals by relaxing the requirement for guaranteed coverage.
  Instead of using separate proportions to construct lower and upper limits, we use a single proportion to construct both.  We call the resulting intervals approximate binomial framework (ABF) CIs.
The ABF CIs will have smaller length CI, but will no longer guarantee coverage.

In this section, assume $G$ is continuous.
Now we develop intervals with approximate coverage using observations on both sides of $t$ to create both confidence limits at once.
In this section, if we are away from the boundaries, then we let $m_n=m(n)$ be a positive even number of observations used to calculate the limits,
with $m_n/2 \leq t$ and $m_n/2 > t$, and we use the closest $m_n/2$  observations to $t$ on either side of $t$.
Close to the boundaries, we modify $m_n$ to keep equal numbers on both sides of $t$, using $m^{\dagger}_n$ observations, where
\begin{equation*}
m^{\dagger}_{n}=2\left[\text{min}\{\lceil m_n/2 \rceil, l_{n}, (n-g_{n}+1)\}\right],
\end{equation*}
where
$l_{n}=\text{max}\{i:C_{i}\leq t\}$,   $g_{n}=\text{min}\{i:C_{i}\geq t\}$, and $m^{\dagger}_{n}\leq n$.
Define
\begin{equation*}
\label{definitionabStarStar}
  a^{\dagger}\equiv a^{\dagger}(t,n,\bold{C})=C_{l_{n}-(m^{\dagger}_{n}/2)+1},~\text{and}~
 b^{\dagger}\equiv b^{\dagger}(t,n,\bold{C})=C_{g_{n}+(m^{\dagger}_{n}/2)-1}.
\end{equation*}
Then if $G$ is continuous, there are ($m^{\dagger}_{n}/2$) observations  in $[a^{\dagger}, t]$ and $[t, b^{\dagger}]$.
The value $m^{\dagger}_{n}$ may be very small at $t$ where $F(t)\approx 0$ or 1. We adjust the confidence interval for this in Section~\ref{sub_sec_CI_adjustment}.

Analogously to  before, we use the form of the Clopper-Pearson two sided $100(1-\alpha)$\% confidence interval functions
(i.e., $L$ and $U$ as in (\ref{defn:LandU})), except now we use $Y_{a^{\dagger}}^{b^{\dagger}}$
and $N_{a^{\dagger}}^{b^{\dagger}} = m^{\dagger}_{n}$.  Specifically,
the $100(1-\alpha)\%$ interval is
\begin{eqnarray}
\left[ L \{1 -\alpha/2; Y^{b^{\dagger}}_{a^{\dagger}}, m^{\dagger}_{n} \}, \right.
& &
\left.
U \{1-(\alpha/2); Y^{b^{\dagger}}_{a^{\dagger}}, m^{\dagger}_{n} \} \right].  \label{CP_surrounding_t}
\end{eqnarray}

\textsc{Theorem 3.} \textit{Under the conditions of Theorem 2.2, the conditional (on ${\bf C}$) and unconditional coverage rates of (\ref{CP_surrounding_t}) tend to $1-\alpha$ as $n\rightarrow \infty$.}\\

The proof is not given since it is very similar to that of Theorem 2.2.

Another adjustment is the mid-p ABF CIs, defined by replacing the functions $L$ and $U$
 in equation~(\ref{CP_surrounding_t})
with the  mid-P binomial confidence limit functions, $L_{mid}$ and $U_{mid}$, defined in \cite[][Section~S.2]{kim2018valid}.
Since the mid-p ABF CIs and usual ABF CIs are asymptotically equivalent,
we work with $L$ and $U$ in the following sections.

\mathversion{bold}
\subsection{Optimal $m^{\dagger}_{n}$ Observations Surrounding $t$ for the Confidence Interval about $F(t)$}
\mathversion{normal}

In Section~\ref{Sec_Opti_Min_Length} we found the $m_n$ that minimized the expected length based on a linear approximation to the
$F(C_i)$ values close to $F(t)$ (see (\ref{approxYat2}) and (\ref{approxYtb2})). Because of the validity requirement, the expected proportion of events was biased since $E(Y_{a}^t/m_n )<F(t)$ and $E(Y_{t}^b/m_n) > F(t)$,
even though the bias decreased with increasing $m_n$. For fixed $n$, increasing $m$  decreases the variance but increases the bias. Thus, we could solve for
minimum expected confidence interval length. For this section, there is no inherent bias, and we cannot solve for minimizing the expected confidence interval length based on the linear
approximation, since that approximation would suggest using $m=n$ to minimize the variance.
Instead we solve for an $m_n$ using two expected mean squared errors (MSEs).

Let the sum of the expected MSEs as a function of $m^{\dagger}_n$ be
\begin{eqnarray*}
Q(m_{n}^{\dagger}) & = & \text{E}\left[ \left\{Y_{a^{\dagger}}^{t}/(m^{\dagger}_{n}/2)-F(t)\right\}^{2}+ \left\{Y_{t}^{b^{\dagger}}/(m^{\dagger}_{n}/2)-F(t)\right\}^{2}\right]
\end{eqnarray*}
and let
\begin{eqnarray*}
m^{\dagger*}_{n} & = & \underset{m^{\dagger}_{n}}{\operatorname{argmin}}~ Q(m_{n}^{\dagger}).
\end{eqnarray*}
Using approximations similar to (\ref{approxYat2}) and (\ref{approxYtb2}),
we approximate  the distributions of $Y_{a^{\dagger}}^{t}$ and $Y_{t}^{b^{\dagger}}$ as
\begin{equation}
\label{approx_M_surrounding}
Y_{a^{\dagger}}^{t}~\dot{\sim}~\text{Binomial}\left[\frac{m^{\dagger}_{n}}{2}, F(t)-\left \{\frac{m^{\dagger}_{n} f(t)}{4ng(t)}\right \} \right];
Y_{t}^{b^{\dagger}}~\dot{\sim}~\text{Binomial}\left[\frac{m^{\dagger}_{n}}{2}, F(t)+\left \{\frac{m^{\dagger}_{n} f(t) }{4ng(t)}\right \} \right].
\end{equation}
This gives the approximation,
\begin{equation}
\label{Qeq}
\begin{array}{l l}
Q(m_{n}^{\dagger})
& \approx 2\left[\frac{2F(t)}{m_{n}^{\dagger}}-\frac{2\{F(t)\}^{2}}{m_{n}^{\dagger}}-2\left(\frac{f(t)}{4ng(t)}\right)^{2}m_{n}^{\dagger}+\left(\frac{f(t)}{4ng(t)}\right)^{2}\left(m_{n}^{\dagger}\right)^{2}\right].
\end{array}
\end{equation}
After taking the derivative of (\ref{Qeq}) with respect to $m_{n}^{\dagger}$ and then setting it to zero, we can find the numerical solution of $m_{n}^{\dagger*}$ from
\begin{equation}
\label{dQeq}
\frac{dQ(m_{n}^{\dagger})}{dm_{n}^{\dagger}}\approx \left\{\frac{f(t)}{4ng(t)}\right\}^{2} (m_{n}^{\dagger})^{3}-\left\{\frac{f(t)}{4ng(t)}\right\}^{2} (m_{n}^{\dagger})^{2}-F(t)+\left\{F(t)\right\}^{2}=0.
\end{equation}
From  Cardano's formula (\cite{cardano1968ars}), we can compute the order of $m_{n}^{\dagger*}$:
\[
\begin{array}{l l}
m_{n}^{\dagger*}& \approx 
\sqrt[3]{\left(\frac{1}{27}+\frac{F(t) - \{F(t)\}^{2}}{2[\{f(t)\}^{2}/\{4ng(t)\}^{2}]}\right)+
\sqrt{\left(\frac{1}{27}+\frac{F(t) - \{F(t)\}^{2}}{2[\{f(t)\}^{2}/\{4ng(t)\}^{2}]}\right)^{2}-
\left(\frac{1}{9}\right)^{3}}}\\[5pt]
&+\sqrt[3]{\left(\frac{1}{27}+\frac{F(t) - \{F(t)\}^{2}}{2[\{f(t)\}^{2}/\{4ng(t)\}^{2}]}\right)-
\sqrt{\left(\frac{1}{27}+\frac{F(t) - \{F(t)\}^{2}}{2[\{f(t)\}^{2}/\{4ng(t)\}^{2}]}\right)^{2}-
\left(\frac{1}{9}\right)^{3}}}+\left(\frac{1}{3}\right)\\[5pt]
&=O(n^{2/3}),
\end{array}
\]
which is a real number.
This method yields a similar conclusion that $m_n$ should be of the order $n^{2/3}$.


 To estimate $m^{\dagger*}_{n}$, we need estimates of  $F(t)$, $f(t)$ and $g(t)$. The value $g(t)$ can be estimated by kernel density estimation with assessment times $C_{i}$, $i=1\ldots n$. But
to estimate $F(t)$ and $f(t)$, we use a slightly modified version of the smoothed maximum likelihood estimation (SMLE)  introduced by \cite{groeneboom2010maximum}. Details are in \cite[][Section~S.3]{kim2018valid}.

\subsection{Confidence Interval with Monotonic Adjustments}
\label{sub_sec_CI_adjustment}

Before describing adjustments for monotonicity, we introduce an additional practical adjustment.
We set the lower confidence limit to 0 when NPMLE $\hat{F}_{n}(t)=0$ and the upper confidence limit to 1 when NPMLE $\hat{F}_{n}(t)=1$.
This adjustment was motivated by preliminary simulations which showed that the edges of the distribution
had poor coverage. Besides leading to better coverage, it ensures that the confidence limits enclose the NPMLE when it reaches those extremes.

Note that we assume that $F(t)$ is  a monotonically increasing  function of $t$.
However the lower and upper limits of the confidence interval  (\ref{CP_surrounding_t})  are not necessarily monotonically increasing  functions of $t$. In this section, we consider two adjustments to construct monotonically increasing lower and upper limits of $F(t)$.

Suppose that the goal is to construct  the monotonically increasing
$k^{\prime}$
pointwise confidence intervals about $F(t_{i})$ where $i=1\ldots k^{\prime}$; $0<t_{1}\leq t_{2} \leq \ldots t_{k^{\prime}}<\infty$.
Let the lower limit and upper limit of the  confidence interval  at $t=t_{i}$  be $L_{F(t_{i})}$ and $U_{F(t_{i})}$, respectively.

 First we consider the edge adjustment. Let  $L_{F(t^{\text{min}}_{\text{L}})}$$=\text{min}\left\{L_{F(t)}\right\}$ for $0<t\leq t_{\lceil k^{\prime}/2\rceil}$, and $L_{F(t^{\text{max}}_{\text{L}})}$$=\text{max}\left\{L_{F(t)}\right\}$ for $t_{\lceil k^{\prime}/2\rceil}\leq t<\infty$. Then set $L_{F(t)}=L_{F(t^{\text{min}}_{\text{L}})}$ for $t\leq t^{\text{min}}_{\text{L}}$, and $L_{F(t)}=L_{F(t^{\text{max}}_{\text{L}})}$ for $t^{\text{max}}_{\text{L}}\leq t$. Analogously, let  $U_{F(t^{\text{min}}_{\text{U}})}$$=\text{min}\left\{U_{F(t)}\right\}$ for $0<t\leq t_{\lceil k^{\prime}/2\rceil}$, and $U_{F(t^{\text{max}}_{\text{U}})}$$=\text{max}\left\{U_{F(t)}\right\}$ for $t_{\lceil k^{\prime}/2\rceil}\leq t<\infty$. Then set $U_{F(t)}=U_{F(t^{\text{min}}_{\text{U}})}$ for $t\leq t^{\text{min}}_{\text{U}}$, and $U_{F(t)}=U_{F(t^{\text{max}}_{\text{U}})}$ for $ t^{\text{max}}_{\text{U}}\leq t$  \citep[see the second panel in Figure~S1][]{kim2018valid}.

Now  consider an adjustment we call the  lower-upper adjustment. For the lower limit, if $t_{i}\leq t_{i+1}$, but  $L_{F(t_{i+1})}\leq L_{F(t_{i})}$ then we replace $L_{F(t_{i+1})}$ with $L_{F(t_{i})}$. We start this lower limit adjustment  from the smallest $t$ and proceed to the largest $t$.
For the upper limit, if $t_{i}\leq t_{i+1}$, but  $U_{F(t_{i+1})}\leq U_{F(t_{i})}$ then we replace $U_{F(t_{i})}$ with $U_{F(t_{i+1})}$. We start this upper limit adjustment  from the largest $t$ and proceed to the smallest $t$.
Then  lower and upper limits of $F(t)$  become monotonically increasing functions, and this adjustment shortens the length of the confidence interval  \citep[see the third panel in Figure S1][]{kim2018valid}.

Now consider the ``middle value'' adjustment.  Let $L^{\text{1}}_{F(t)}$ be the lower limit function from the lower-upper adjustment just described.
Let $L^{\text{2}}_{F(t)}$, be similar except we start at the opposite end.
As before,
 if $t_{i}\leq t_{i+1}$, but  $L_{F(t_{i+1})}\leq L_{F(t_{i})}$ then $L^{\text{2}}_{F(t_{i+1})}$ is defined as $L_{F(t_{i})}$, but we start this adjustment  from the largest $t$ proceeding to the smallest $t$.
 Then the lower limit with the middle value adjustment  is $L^{\text{M}}_{F(t)}=\{L^{\text{1}}_{F(t)}+L^{\text{2}}_{F(t)}\}/2$.
Analogously, we define $U^{\text{1}}_{F(t)}$ as an upper limit ensuring monotonicity by proceeding from the smallest to the largest $t$,
and define $U^{\text{2}}_{F(t)}$ as the upper limit ensuring monotonicity by proceeding from the largest to the smallest $t$.
Then the upper limit with the middle value adjustment  is $U^{\text{M}}_{F(t)}=\{U^{\text{1}}_{F(t)}+U^{\text{2}}_{F(t)}\}/2$  \citep[see the fourth panel in Figure S1][]{kim2018valid}. Then  lower and upper limits of $F(t)$ with the middle value adjustment  are monotonically increasing functions.


\section{Simulation Studies}
\label{sectionSimulation}

\subsection{Simulation 1}
In this section, we perform  simulation studies.
We begin with a simulation described as Case 1 ($f(t)$ is  Exp(1), and $g(t)$ is  Exp(1)  for $0\leq t<\infty$) with $n=1,000$,
and using confidence interval methods described in \cite[Section 9.5]{groeneboom2014nonparametric}.
Specifically, we set $a=0$ and $b$ as the maximum of the assessment times (see their equation 9.75),
We used the triweight kernel and set the bandwidth to $h=F^{-1}(0.99) n^{-1/4},$ where here $F^{-1}(0.99)=4.605$ which comes from the true distribution, so $h\approx \text{(Range of the assessment times)}\times n^{-1/4}$.
We generated $1,000$ bootstrap samples, and for the 95\% confidence interval we used the $20$th and $980$th of the bootstrap samples
\citep[to adjust for undercoverage, see][p. 272]{groeneboom2014nonparametric}.
Despite these adjustments, there was substantial undercoverage \citep[see][Figure S2]{kim2018valid}.
There could be other choices for how to implement those confidence intervals, but since this implementation did not perform well, we
do not include these methods in the full simulation results.

For the full simulation,
we consider three possible cases:
\begin{enumerate}[label=\bfseries Case \arabic*:]
  \item $f(t)$ is  Exp(1), $f(t)$=$exp(-t)$, and $g(t)$ is  Exp(1), $g(t)$=$\exp(-t)$ for $0\leq t<\infty$;
  \item $f(t)$ is  Gamma(3,1), $f(t)$=$[1/\{3\Gamma(1)\}] \exp\{-(t/3)\}$ for $0\leq t<\infty$, and $g(t)$ is  Unif(0,5), $g(t)$=$t/5$ for $0\leq t\leq 5$;
  \item $f(t)$ is the mixure of Gamma(3,1) and Weibull(8,10), \\
$f(t)$=$.5[1/\{3\Gamma(1)\}]$$\exp\{-(t/3)\}$$+.5\{(8/10)(t/10)^{7}\}$$\exp\{-(t/10)^{8}\}$ for $0\leq t<\infty$, and $g(t)$ is  Unif(0,15), $g(t)$=$t/15$ for $0\leq t\leq 15$.
\end{enumerate}
Then  the  two-sided 95\% CIs (\ref{CP_surrounding_t}) have been constructed about $0<F(t)<1$.
We use the likelihood ratio-based CI introduced by \cite{banerjee2001likelihood}  as a benchmark, to compare to our new methods.

\begin{figure}[t]
\centering
\includegraphics[height=5in,width=5in]{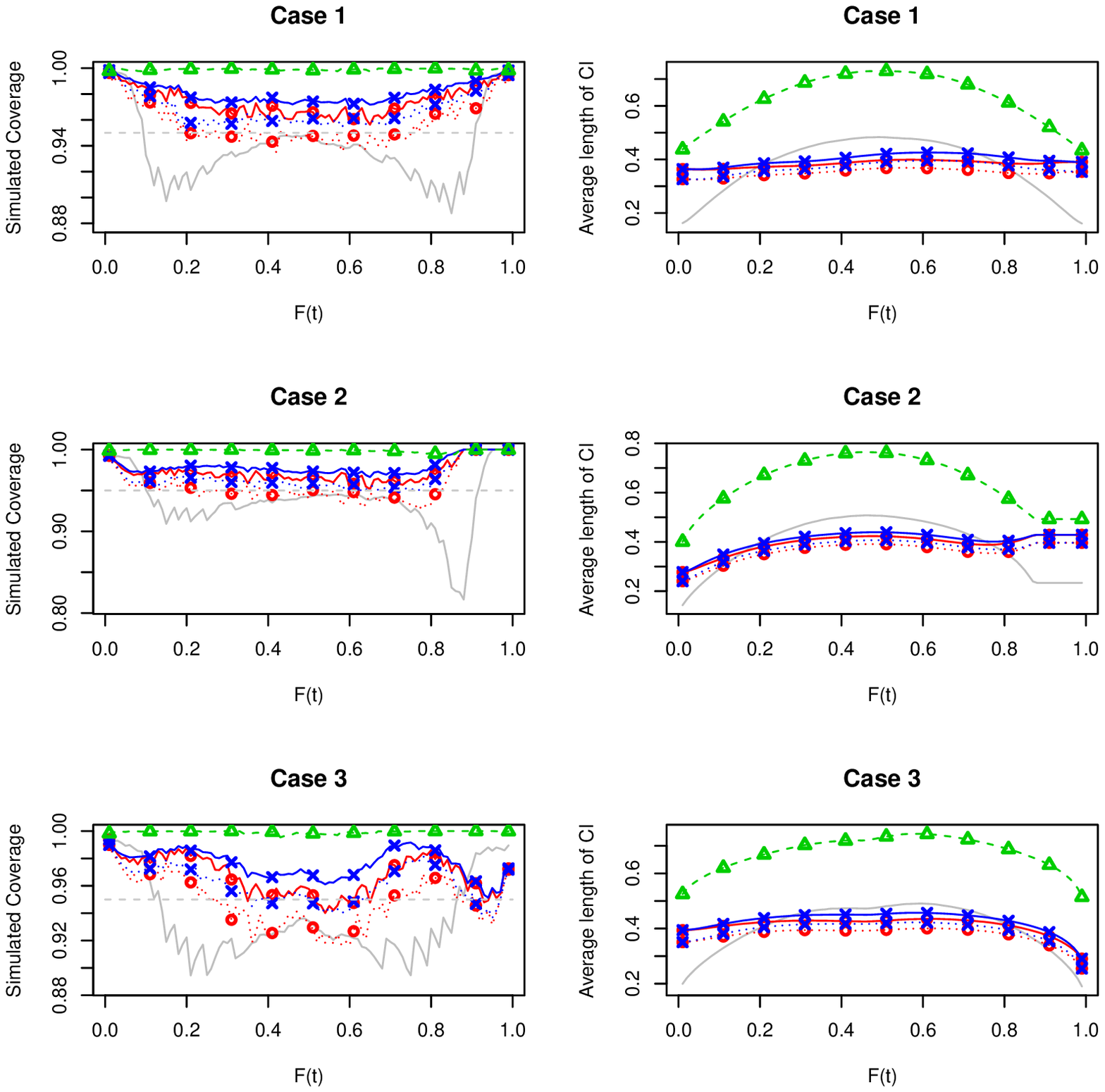}
 \caption{\textit{Simulated coverage and average length of 95 \% confidence intervals for Case 1,2, and 3. The likelihood ratio-based CI (gray solid line);
 the ABF CI with edge and lower-upper adjustment (red solid line with ($\circ$));
 the mid-$P$ ABF CI with edge and lower-upper adjustment (red dotted line with ($\circ$));
 the ABF CI with edge and middle value adjustment (blue solid line with ($\times$));
 the mid-$P$ ABF CI with edge and middle value adjustment (blue dotted line with ($\times$));
 the valid CI  with $m_{n}=n^{2/3}$ (green dashed line with ($\triangle$)).  The sample size is $n$=50, and 10,000 replications have been performed.}}
  \label{Figure_Six_Plots}
\end{figure}

In Figure \ref{Figure_Six_Plots}, we plot the simulated coverage and the averaged lengths of the six different CIs for $0< F(t)<1$:  the likelihood ratio-based CI, the ABF CI with edge \& lower-upper adjustment, the mid-$P$ ABF CI with edge \& lower-upper adjustment, the ABF CI with edge \& middle value adjustment, the mid-$P$ ABF CI with edge \& middle value adjustment, and the valid CI  with  $m_n= n^{2/3}$.
Each simulation used 10,000 replications and had  $n=50$. The cases $n=200$ and $n=1,000$ are plotted  in Figures~S3 and S4 of \cite{kim2018valid}.

Figures show that the ABF CIs   have generally shorter length than the likelihood ratio-based CIs, but have better coverage. When the likelihood ratio-based CIs have adequate coverage, the ABF CIs have shorter length.
The ABF CIs seem to be conservative with  small $n$ (e.g. $n=50$), but this conservativeness can be eliminated by using the mid-$P$ approach.

Since the lower-upper adjustment shortens the length of the CI,
the ABF CI with the lower-upper adjustment has relatively shorter length than the ABF CI with edge and middle value adjustment. We do not recommend  using the mid-$P$ approach with the lower-upper adjustment simultaneously,
since that combination doubly shortens the length of the CI to such a degree that the coverage is poor (see Case 3 in Figure \ref{Figure_Six_Plots}).
 Therefore, we recommend using either the ABF CI with edge and lower-upper adjustment or the mid-$P$ ABF CI with  edge and  middle value adjustment.

\subsection{Simulation 2}
\begin{figure}[t]
        \centering
        \includegraphics[height=5in,width=5in]{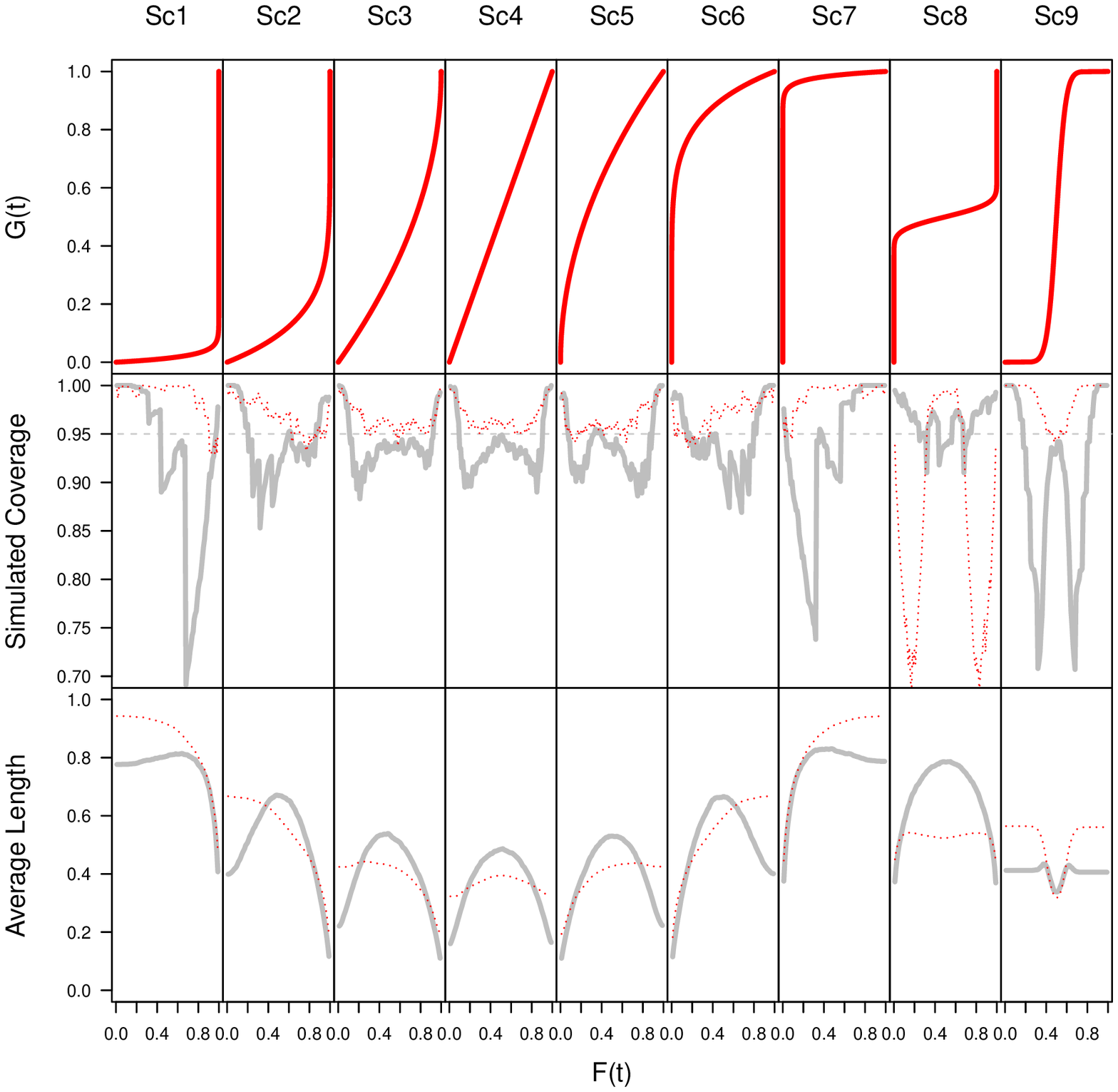}
    \caption{\textit{Simulated coverage and average length of 95 \% CIs  : the likelihood ratio-based CI (gray solid line); the mid-$P$ ABF CI with edge and middle value adjustment (red dotted line) with $n$ = 50. There are 9 scenarios which are described in text, and simulations
based on 1,000 replications.}}
\label{Scenario9_n50_Plots}
\end{figure}

In this section, we consider more extensive and systematic simulations. We assume nine possible scenarios assuming that $g(t)\sim\text{Unif}(0,1)$ and $f(t)\sim \text{Beta}(\alpha,\beta)$ where $\alpha=1, \beta=50$; $\alpha=1, \beta=7$; $\alpha=1, \beta=2$; $\alpha=1, \beta=1$; $\alpha=2, \beta=1$; $\alpha=7, \beta=1$; $\alpha=50, \beta=1$; $\alpha=100, \beta=100$; and $\alpha=.1, \beta=.1$.  Figure \ref{Scenario9_n50_Plots} shows simulated coverage and averaged lengths of 95 \% CIs  based on the likelihood ratio-based CI and the mid-$P$ ABF CI with edge and middle value adjustment when $n$ = 50. The ABF CIs  have generally shorter length than the likelihood ratio-based CIs, but have better coverage. When the function $F(t)$ rises very steeply (Scenario 8), the ABF CI has poor coverage at the areas of big changes of the slope. However, the coverage approaches the nominal rate as $n$ becomes larger.
Figure S5 and S6 in \cite{kim2018valid} show simulated coverage and average lengths of 95 \% CIs when $n=200$ and $n=1,000$.

 \cite{banerjee2005confidence}
considered a  setting when $t$ lies in a region of steep ascent of the distribution function $F(t)$. They assumed $g\sim \text{Unif}(0,1)$ and

\begin{equation*}
F(t)= \left\{
  \begin{array}{l l}
    t& \quad \text{for $t\leq .25$ };\\
    .25+(20,000)(t-.25)^2 & \quad \text{for $.25<t\leq \left(.25+\frac{1}{200}\right)$};\\
    .75+\frac{.25}{(.75-\frac{1}{200})}(t-.25-\frac{1}{200}) & \quad \text{for $\left(.25+\frac{1}{200}\right)<t\leq 1$}.
  \end{array} \right.
  \end{equation*}
This case is similar to Scenario 8. However, in this case, there are  two points with discontinuous derivatives: when $F(t)=.25$ and $F(t)=.75$.  Figure S7 in \cite{kim2018valid} shows simulated coverage and average lengths of 95 \% CIs when $n$=50, 200, and 1,000. The ABF CI has very poor coverage around the points with discontinuous derivatives. Even when $n$ becomes larger, the coverage is still poor around those points. However the ABF CI has good coverage at the edges, and in the center.  This simulation setting tells us that the ABF CI can perform poorly when in areas where $F(t)$ changes very dramatically with possibly discontinuous derivatives.

\section{Analyzing the hepatitis A data in Bulgaria}
\label{sectionApplication}

\begin{figure}[h]
\centering
\includegraphics[height=5in,width=5in]{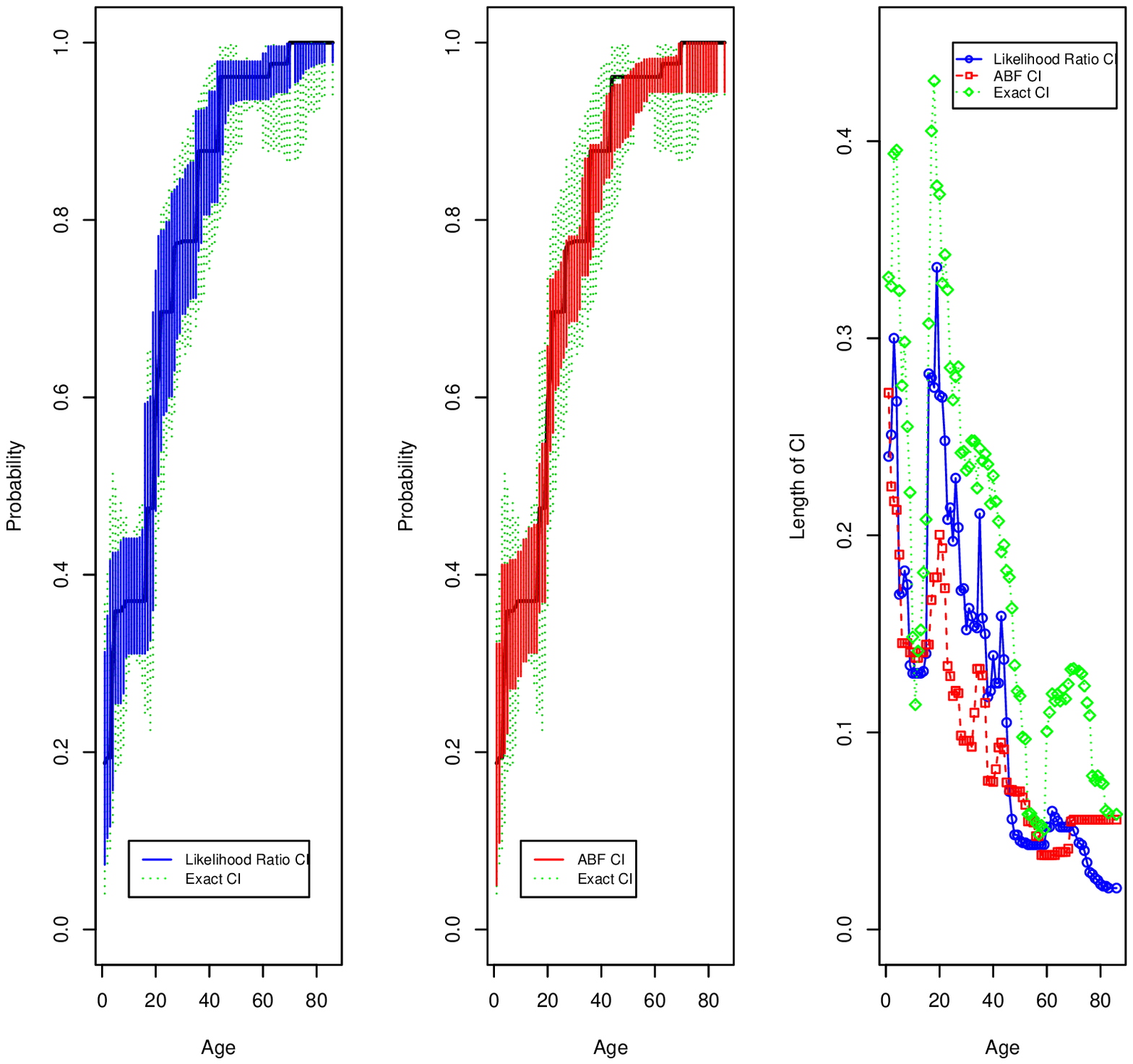}
 \caption{\small \textit{Hepatitis A data: 95\% confidence intervals for $F(t)$, the probability of ever being infected prior to or at age $t$.
 Left panel: the likelihood ratio-based CIs;  middle panel: the mid-$P$ ABF CIs  with edge and middle value adjustment. The dotted vertical lines are valid CIs with $m=n^{2/3}$,
 and  the solid step functions inside the confidence intervals are the NPMLE  in the left and the middle panels.  Right panel: comparison of  confidence interval lengths;  the likelihood ratio-based CI (blue solid line with ($\circ$));  the mid-$P$ ABF CIs  with edge and middle value adjustment (red dashed line with ($\Box$)); the valid CIs with $m=n^{2/3}$
  (green dotted line with ($\diamond$)).  }}
  \label{Figure_LR_OMSM_Exact}
\end{figure}

\cite{keiding1991age} analyzed data on anti-hepatitis A antibody responses in Bulgaria.
For the purpose of this analysis, we assume that once a person has been infected with hepatitis A,
that person will test positive for anti-hepatitis A antibodies throughout the remainder of his or her life.
Further, we assume that the force of infection of hepatitis A does not change over time.
Thus, a cross-sectional sample can be interpreted as current status data, where
the time scale is age and the event is a positive test for  anti-hepatitis A antibodies.
  Table 2 in \cite{keiding1991age} contains the data, consisting of
   $850$ people whose ages range from 1-86 years.
At each single year age group we have the number of people tested and the number of those who tested positive (a few ages had no one tested).
The main goal here is to construct the pointwise confidence intervals of  the distribution of age at which people were first exposed to hepatitis A.
Based on this current status data, we constructed the likelihood ratio-based CI, the mid-$P$ ABF CI with edge and middle value adjustment, and the valid CI  with  $m_n = n^{2/3}$.
In Figure \ref{Figure_LR_OMSM_Exact}, the mid-$P$ ABF CIs are seen to be  shorter  than the likelihood ratio-based CIs in the middle of the age range.  The mid-$P$ ABF CIs are slightly wider than the likelihood ratio-based CIs at the right edge (especially, when the NPMLE, $\hat{F}(t)=1$).  Some mid-$P$ ABF CIs with edge and middle value adjustment do not contain NPMLE values (see the middle panel in Figure \ref{Figure_LR_OMSM_Exact}). This may happen for some points of $t$  because the ABF CI is based on the local binomial-type responses, not the  NPMLE.

\section{CONCLUSION}
\label{conclusion}

We introduced a new framework for CI with current status data.
We developed two new types of CIs: the valid CI and the ABF CI. The valid CI  guarantees the nominal coverage rate
and approaches the nominal rate if $m_{n}$ satisfies the asymptotic conditions.
 The valid method is simple and can be applied with continuous or discrete assessment distributions.
 The ABF CI does not guarantee the nominal rate, but its coverage rate  asymptotically approaches the nominal rate  if $m^{\dagger}_{n}$ satisfies the asymptotic conditions. In a series of simulations, we compare our new CIs to the LRT CI, and no one method outperforms the others in every situation.
 When guaranteed coverage is needed, then the valid CIs
are recommended. When an approximation with shorter confidence interval lengths is acceptable, then either
the LRT CI or the ABF CI are appropriate.  When the failure time distribution is changing rapidly in an area where there is not a high density of assessments,
 then the LRT CI often has coverage closer to the nominal than the ABF CI; however, in most other cases the ABF CI showed better coverage than the LRT CI, especially in the areas away from the middle of the distribution.

\clearpage

\setcounter{section}{0}
\renewcommand{\thesection}{\Alph{section}}

\appendix

\section{PROOFS}\label{app}
\subsection{Proof of Equation \ref{lowerfixa}}
\label{Appendix_equation_3}

Given $C_{i}=c_i$,
\[
D_{i}|C_{i}=c_i~\sim~\text{Bernoulli}~\{ F(c_{i})\} .
\]
Let $D_{i}^{*}$ be a random variable such that
$D_{i}^{*}~\sim~\text{Bernoulli}~\{ F(t)\}$. The $C_i$ used for the lower bound for $F(t)$ are $\le t$, so $F(c_i)\le F(t)$.
It is clear that a Bernoulli$(p)$ random variable becomes stochastically larger as $p$ increases, so $D_i\,|\,C_i=c_i$ is less than or equal
to a Bernoulli with parameter $F(t)$ in the stochastic order (\cite{whitt2006stochastic}), which we write $D_i\,|\,C_i=c_i  \preceq D_{i}^{*}$.

Since $D_{i}$ and $D^{*}_{i}$ are independent sets of random variables with $D_{i} \preceq D_{i}^{*}$ for each $i$, $Y(a,t)|\bold{C}=\sum_{i:C_{i}\in[a,t]} D_{i}|C_{i}$ $\preceq$ $B|\bold{C}$ $\sim$ Binomial\{$N(a,t)$, $F(t)$\} (see theorem 1.A.3, part b, of
\cite{shaked2007stochastic}).

Let
\begin{equation*}
\begin{array}{l}
g(y)=L(q;y,n)= \left\{
  \begin{array}{l l}
    0& \quad \text{if  $y=0$}\\
    Be\{1-q; y, n-y+1\} & \quad  \text{if $y>0$}
  \end{array} \right.
  \end{array}
\end{equation*}
be the lower $q$th one-sided Clopper-Pearson confidence interval. Since $L(q;y,n)$ is a monotonic increasing function of $y$ given fixed $q$ and $n$,
$g(Y(a,t)|\bold{c})$ $\preceq$ $g(B|\bold{c})$ outside a null set of ${\bf c}$ values.
Therefore,  for given $q$ and $N(a,t)$, outside a null set of ${\bf c}$ values,
\[
\text{Pr}[L\{q; Y(a,t), N(a,t)\}\leq F(t)|\bold{c}]\geq \text{Pr}[L\{q; B, N(a,t)\}\leq F(t)|\bold{c}] \geq q.
\]
Now take the expectation of each side over the distribution of ${\bf C}$ to conclude that
\[
\text{Pr}[L\{q; Y(a,t), N(a,t)\}\leq F(t)] \geq q. ~~~\Box
\]

\subsection{Proof of Theorem 1a}
\label{Appendix_Proof_1a}

This follows by conditioning on ${\bold C}$ and noting that (\ref{lowerfixa}) and (\ref{upperfixb}) include conditional probabilities given ${\bold C}$.  With probability $1$, the conditional probability that either $L>F(t)$ or $U<F(t)$ is no greater than the sum of these probabilities, which is no greater than $\alpha/2+\alpha/2=\alpha$.  Because the conditional coverage probability is at least $1-\alpha$ (almost surely), the unconditional probability, namely the expectation of the conditional probability, is at least $1-\alpha$ as well. Centrality follows because each of the error probabilities is less than or equal to $\alpha/2$.
$\Box$

\subsection{Proof of Theorem 2.1}
\label{Appendix_Proof_2a}

First consider the coverage rate of (\ref{twoside}).
Let $C_1^\prime,C_2^\prime,\ldots$ be  independent and identically distributed from $G$ (i.e., they are the unordered assessment times).
Let $c^\prime$ be an atom of $G$, and let $p_{c^\prime}>0$ be its probability.  By the strong law of large numbers, the sample proportion of  $C_1^\prime,\ldots,C_n^\prime$ that equal $c^\prime$ converges to $p_{c^\prime}$ with probability $1$.  Consider an $\omega$ for which this happens.  Because $m_n/n\rightarrow 0$ by assumption, the number of $C_1^\prime,\ldots,C_n^\prime$ equaling $c^\prime$ will exceed $m_n$ for all but a finite number of $n$.  Therefore, if $t=c'$ then $a_n$ will equal $t$ for all but finitely many $n$.
Let $D_i^\prime$ be the $D_j$ associated with $C_i^\prime$. Then there are at least $m_n$  $D_i^\prime$ values with $C_i^\prime=t$ corresponding to iid Bernoullis with probability $F(t)$.  The lower limit is the same as the one based on the empirical distribution function from a sample of at least $m_n$  iid observations from distribution $F$, and similarly for the upper limit.  It is known that the coverage probability for a confidence interval based on an empirical distribution function converges to $1-\alpha$ as the sample size tends to $\infty$.

We have shown that, with probability $1$, the conditional coverage probability in (\ref{twoside}) tends to $1-\alpha$ as $n\rightarrow \infty$ at each atom of $G$.  The unconditional coverage probability, namely the expectation of the conditional coverage probability, also tends to $1-\alpha$ at each atom of $G$ by the bounded convergence theorem.

Now consider the coverage rate of   (\ref{twoside_with_m}).
We have already shown that, except for finitely many $n$, the lower and upper intervals correspond to those using the empirical distribution function of a sample of $m_n$ from $F$, in which case the lower limit cannot exceed the upper limit.  Therefore, the coverage probability of the modified interval $[L^*,U^*]$ tends to $1-\alpha$ as $n\rightarrow \infty$ as well.
$\Box$

\subsection{Proof of Theorem 2.2}
\label{Appendix_Proof_2b_i}

In our notation, $C_1\le C_2\le\ldots\le C_n$ are ordered and $T_i$ is the survival time associated with $C_i$.  It is also convenient to think instead of the
infinite set of iid pairs $(C_1^\prime,T_1^\prime),(C_2^\prime,T_2^\prime),\ldots$, where the $C_i^\prime$ are unordered.  Thus, $C_1,\ldots,C_n$ are the order statistics of $C_1^\prime,\ldots,C_n^\prime$, and $T_i^\prime$ is the survival time associated with the $i$th unordered assessment time $C_i^\prime$.
Assume the following conditions:
\begin{equation}
F \mbox{and } G \mbox{ are continuous on $\Re^+$  }  \mbox{ and } F^\prime(t)=f(t)>0,\ G^\prime(t)=g(t)>0. \label{condFandG}
\end{equation}
Note that (\ref{condFandG}) concerns the derivative of $F$ and $G$ at the single point $t$.

Go backwards in time from $t$ to find the $1$st $C^\prime$ (the one closest to $t$ among those less than $t$), $2$nd $C^\prime$ (the one second closest to $t$ among those less than $t$),$\ldots,m_n$th $C^\prime$. Let $a_n$ denote the $m_n$th such point, or $0$ if there are fewer than $m_n$ $C^\prime$s less than $t$.  Note that $a_n$ is a function of the $C_i^\prime$s, hence is a random variable, but $m_n$ is nonrandom.  Assume the following:
\begin{equation}
m_n\rightarrow \infty,\ \ m_n/n\rightarrow 0.  \label{initassum}
\end{equation}
These conditions imply
\begin{equation}
a_n\rightarrow t \mbox{ almost surely.} \label{angoestot}
\end{equation}

We are interested in conditional distribution functions given assessment times.  Let
${\cal C}^\prime_\infty=\{C_1^\prime,C_2^\prime,\ldots\}$, and let ${\cal C}^\prime_{[a_n,t]}$ be the collection of $C_1^\prime,C_2^\prime,\ldots,C_n^\prime$ that are in the interval $[a_n,t]$.  Given ${\cal C}^\prime_\infty$, the sum $Y_{a_n}^{t}$ of indicators of death by the times ${\cal C}^\prime_{[a_n,t]}$ are independent Bernoulli's with probability parameters $F(c_i^\prime)$, where each $c_i^\prime$ is the realized value of $C_i^\prime$ among those in ${\cal C}^\prime_{[a_n,t]}$.
With probability $1$, the number of those independent Bernoulli's is $m_n$ for all sufficiently large $n$.
 We try to approximate the conditional distribution of $Y_{a_n}^{t}$ given ${\cal C}^\prime_\infty$ by a binomial with parameters $\{m_n,F(t)\}$.

Note that, given ${\cal C}^\prime_\infty$, $Y_{a_n}^{t}$ is stochastically larger than or equal to a sum of $m_n$ iid Bernoullis with probability parameter $F(a_n)$, and stochastically smaller than or equal to a sum of $m_n$ iid Bernoullis with parameter $F(t)$.  That is, given ${\cal C}^\prime_\infty$, $Y_{a_n}^{t}$ is stochastically between a binomial random variable $W_{m_n}^\prime={\rm bin}\{m_n,F(a_n)\}$ and a binomial random variable $W_n={\rm bin}\{m_n,F(t)\}$.  Therefore, we seek necessary and sufficient conditions for $m_n$ such that the conditional distributions of
\begin{equation}
Z_n={W_{m_n}-m_nF(t) \over \sqrt{m_nF(t)\{1-F(t)\}}}
 \mbox{  and  }
 Z_n^\prime={W_{m_n}^\prime-m_nF(t) \over \sqrt{m_nF(t)\{1-F(t)\}}}
 \label{ZnZnprime}
\end{equation}
given ${\cal C}^\prime_\infty$ both converge to standard normals as $n\rightarrow \infty$.  The result for $Z_n$ follows immediately from the ordinary CLT, so we need only find necessary and sufficient conditions under which the result holds for $Z_n^\prime$.

Write
\begin{scriptsize}
\begin{eqnarray}
Z_n^\prime&=&{W_{m_n}^\prime-m_nF(a_n) \over \sqrt{m_nF(a_n)\{1-F(a_n)\}}}\sqrt{{F(a_n)\{1-F(a_n)\}\over F(t)\{1-F(t)\}}}+{m_n\{F(a_n)-F(t)\}\over \sqrt{m_nF(t)\{1-F(t)\}}} \cr
   & & \cr
   & & \cr
   &=&{W_{m_n}^\prime-m_nF(a_n) \over \sqrt{m_nF(a_n)\{1-F(a_n)\}}}\sqrt{{F(a_n)\{1-F(a_n)\}\over F(t)\{1-F(t)\}}}+{\sqrt{m_n}\{F(a_n)-F(t)\}\over \sqrt{F(t)\{1-F(t)\}}} \label{Slutskystep}
\end{eqnarray}
\end{scriptsize}
Conditioned on ${\cal C}^\prime_\infty$, the $a_n$ are fixed constants.  The conditional characteristic function of $Z_n^\prime$ given $C_1^\prime=c_1^\prime,C_2^\prime=c_2^\prime,\ldots$ is
$$
\psi_n(t)\exp\left\{it{\sqrt{m_n}\{F(a_n)-F(t)\}\over \sqrt{F(t)\{1-F(t)\}}}\right\},
$$
where $\psi_n(t)$ is the characteristic function of the left term of (\ref{Slutskystep}). By the initial assumptions (\ref{initassum}), the set of $\omega$ for which
$m_nF\{a_n(\omega)\}[1-F\{a_n(\omega)\}]\rightarrow \infty$ has probability $1$.  For each such $\omega$, the conditional distribution of
$$
{W_{m_n}^\prime-m_nF(a_n) \over \sqrt{m_nF(a_n)\{1-F(a_n)\}}}
$$
given $C_1^\prime=c_1^\prime,C_2^\prime=c_2^\prime,\ldots$ satisfies the Lindeberg condition, and therefore converges in distribution to a standard normal
\citep[see Example 8.15 of][]{Pros:2016}.
This, coupled with (\ref{angoestot}) and Slutsky's theorem, implies that the conditional distribution of the left term of (\ref{Slutskystep}), given $C_1^\prime=c_1^\prime,C_2^\prime=c_2^\prime,\ldots$ converges to a standard normal.  Accordingly, its conditional characteristic function $\psi_n(t)$ converges to $\exp(-t^2/2)$ as $n\rightarrow \infty$.  The conditional characteristic function of $Z_n^\prime$ given $C_1^\prime=c_1^\prime,C_2^\prime=c_2^\prime,\ldots$ converges to $\exp(-t^2/2)$ (namely that of a standard normal) if and only if
$$
\exp\left\{it{\sqrt{m_n}\{F(a_n)-F(t)\}\over \sqrt{F(t)\{1-F(t)\}}}\right\}\rightarrow 1,
$$
which occurs if and only if $m_n^{1/2}\{F(a_n)-F(t)\}\rightarrow 0$.  But
$$
\sqrt{m_n}\{F(a_n)-F(t)\}=\left({F(a_n)-F(t) \over a_n-t}\right) \sqrt{m_n}(a_n-t),
$$
and $\{F(a_n)-F(t)\}/(a_n-t)\rightarrow f(t)>0$ by assumption (\ref{condFandG}).  Therefore, $m_n^{1/2}\{F(a_n)-F(t)\}\rightarrow 0$ if and only if $m_n^{1/2}(t-a_n)\rightarrow 0$.

We have shown that, under assumptions (\ref{condFandG}) and (\ref{initassum}), the conditional distributions of $Z_n^\prime$ and $Z_n$ given $C_1^\prime=c_1^\prime,C_2^\prime=c_2^\prime,\ldots$ both converge to standard normals as $n\rightarrow \infty$ if and only if $m_n^{1/2}(a_n-t)$ converges almost surely to $0$.

We show next that $m_n^{1/2}(a_n-t)$ converges almost surely to $0$ if and only if $m_n/n^{2/3}\rightarrow 0$ as $n\rightarrow \infty$.  Assume first that $m_n/n^{2/3}\rightarrow 0$.  The same argument as above, but applied to $G$ instead of $F$, shows that under conditions (\ref{condFandG}) and (\ref{initassum}), $m_n^{1/2}(a_n-t)\rightarrow 0$ if and only if $m_n^{1/2}\{G(a_n)-G(t)\}\rightarrow 0$.  We will, therefore, demonstrate that $m_n^{1/2}\{G(a_n)-G(t)\}\rightarrow 0$ almost surely.

It suffices to show that
\begin{equation}
P[m_n^{1/2}\{G(t)-G(a_n)\}>\epsilon \mbox{ i.o.}]=0  \label{Gio}
\end{equation}
for each $\epsilon>0$ (where i.o. means infinitely often, i.e., for infinitely many $n$).   By the Borel Cantelli lemma, we need only show that
\begin{equation}
\sum_{n=1}^\infty P(E_n)<\infty, \label{BorelCantelli}
\end{equation}
where
\begin{equation}
E_n\mbox{ is the event that }  m_n^{1/2}\{G(t)-G(a_n)\}>\epsilon.  \label{Endef}
\end{equation}
Note that $E_n$ occurs if and only if fewer than $m_n$ of $G(C_1^\prime),G(C_2^\prime),\ldots,G(C_n^\prime)$ lie in the interval $[G(t)-\epsilon/m_n^{1/2},G(t)]$.  Each $G(C_i^\prime)$ follows a uniform distribution, so the number $N_n$ of $G(C_1^\prime),\ldots,G(C_n^\prime)$ in the interval $[G(t)-\epsilon/m_n^{1/2},G(t)]$
has a binomial distribution with parameters $(n,\epsilon/m_n^{1/2})$.

\cite{Slud:1977} shows that for a binomial $(n,p)$ random variable $X$, if $x\le np$, then $P(X\ge x)\ge 1-\Phi[(x-np)/\{np(1-p)\}^{1/2}]$.  Equivalently,
$P(X<x)\le \Phi[(x-np)/\{np(1-p)\}^{1/2}]$.  We can apply this to conclude that when $m_n<n\epsilon/m_n^{1/2}$ (which it will be for large $n$ because $m_n/n^{2/3}\rightarrow 0$),
$$
P(N_n<m_n)\le \Phi\left[{m_n-{n\epsilon\over \sqrt{m_n}}\over \sqrt{n\left({\epsilon\over \sqrt{m_n}}\right)\left(1-{\epsilon\over \sqrt{n}}\right)}}\right].
$$

It is known that $1-\Phi(x)\le \phi(x)/x$ for $x\ge 0$ \citep[see section 11.11.2 of][]{Pros:2016}, so by symmetry, $\Phi(x)\le \phi(x)/|x|$ for $x\le 0$ as well.  In our case, $x=x_n$, where
$$
x_n={m_n-{n\epsilon\over \sqrt{m_n}}\over \sqrt{n\left({\epsilon\over \sqrt{m_n}}\right)\left(1-{\epsilon\over \sqrt{m_n}}\right)}}.
$$
Therefore, it suffices to show that $\sum_{n=1}^\infty \phi(x_n)/|x_n|<\infty$.  But if $n\rightarrow \infty$ and $m_n/n^{2/3}\rightarrow 0$, then $|x_n|\rightarrow \infty$.  Accordingly, we can ignore the denominator of $\phi(x_n)/|x_n|$ and show that $\sum_{n=1}^\infty \phi(x_n)<\infty$.

To show that
$$
\sum_{n=1}^\infty{\phi\left[{m_n-{n\epsilon\over \sqrt{m_n}}\over \sqrt{n\left({\epsilon\over \sqrt{m_n}}\right)\left(1-{\epsilon\over \sqrt{m_n}}\right)}}\right]}<\infty,
$$
write the $n$th term $d_n$ of the sum as

\begin{align}
d_n&=&(2\pi)^{-1/2}\exp\left\{{-{n^2\over 2m_n}\left(\epsilon-{m_n^{3/2}\over n}\right)^2 \over {n\epsilon\over \sqrt{m_n}}\left(1-{\epsilon\over \sqrt{m_n}}\right)}\right\}=(2\pi)^{-1/2}\exp\left\{{-{n\over 2\sqrt{m_n}}\left(\epsilon-{m_n^{3/2}\over n}\right)^2 \over \epsilon\left(1-{\epsilon\over \sqrt{m_n}}\right)}\right\} \cr
 & &=(2\pi)^{-1/2} \exp\left\{{-(1/2)\left({n^{2/3}\over m_n}\right)\sqrt{m_n}n^{1/3}\left(\epsilon-{m_n^{3/2}\over n}\right)^2 \over \epsilon\left(1-{\epsilon\over \sqrt{m_n}}\right)}\right\}.
\end{align}
The denominator inside the exponent is no greater than $\epsilon$, while $n^{2/3}/m_n\rightarrow \infty$, $m_n^{1/2}\rightarrow\infty$, and $(\epsilon-m_n^{3/2}/n)\rightarrow \epsilon$ as $n\rightarrow \infty$.  It follows that what is inside the exponent is at most
$$
\exp(-\lambda n^{1/3})
$$
for all $n$ sufficiently large, where $\lambda>0$.  Now apply the integral test for infinite sums to conclude that $\sum_{n=1}^\infty \exp(-\lambda n^{1/3})<\infty$ because $\int_1^\infty\exp(-\lambda x^{1/3})dx<\infty$.  We have demonstrated condition (\ref{BorelCantelli}). By the Borel Cantelli lemma, (\ref{Gio}) holds.

We have shown that
\begin{eqnarray}
m_n/n^{2/3}\rightarrow 0&\Rightarrow&  m_n^{1/2}\{G(a_n)-G(t)\}\rightarrow 0 \mbox{ a.s} \cr
                        &\Rightarrow& m_n^{1/2}(a_n-t)\rightarrow 0 \mbox{ a.s} \cr
                        &\Rightarrow& \mbox{given }{\cal C}^\prime_\infty, \mbox{ the conditional distribution functions of }Z_n \mbox{ and } Z_n \cr
                        &           &\mbox{of } (\ref{ZnZnprime}) \mbox{ both converge to standard normals a.s.}
\end{eqnarray}

For the reverse direction, suppose that $m_n/n^{2/3}$ does not converge to $0$.  Then there is some subsequence $\{K\}\subset \{1,2,\ldots\}$ such that $m_k/k^{2/3}$ converges to either a positive number or $+\infty$ for $k\in \{K\},\ k\rightarrow \infty$.  We shall show that in either case, (\ref{Gio}) cannot hold.  With $E_n$ defined by (\ref{Endef}), along the subsequence $\{K\}$, we have
\begin{eqnarray}
P(E_k \mbox{ i.o.})&=&P\left[\cap_{j\in\{K\}} \cup_{r\ge j,\,r\in \{K\}} E_r)\right] \cr
                   &=&\lim_{j\rightarrow \infty,\,j\in \{K\}} P(\cup_{r\ge j,\,r\in \{K\}}E_r) \cr
                   &\ge & \liminf_{j\rightarrow \infty,\,j\in\{K\}} P(E_j) \label{liminfy}
\end{eqnarray}
Also, $m_j/j^{2/3}$ converges to a positive constant or $+\infty$ as $j\rightarrow \infty$ along the subsequence $\{K\}$, so $m_j^{3/2}/j$ converges to $B$, where $B$ is a positive constant or $+\infty$.  This implies that for $\epsilon<B$,  $m_j-j\epsilon/m_j^{1/2}\ge 0$ for all sufficiently large $j$. Consequently, for $\epsilon<B$, $P(E_j)\ge 1/2$ for all sufficiently large $j$. By inequality (\ref{liminfy}), $P(E_k\ \mbox{i.o.})\ge 1/2$. This certainly precludes (\ref{Gio}).  This completes the proof that if $m_n/n^{2/3}$ does not converge to $0$, the conditional distribution of $Z_n$ and $Z_n$ given ${\cal C}^\prime_\infty$ cannot both converge to standard normals almost surely as $n\rightarrow \infty$.

We have shown that the conditional distributions of $Z_n^\prime$ and $Z_n$ given ${\cal C}^\prime_\infty$ both converge to standard normals a.s. as $n\rightarrow \infty$ if and only if $m_n/n^{2/3}\rightarrow 0$.

\section*{Acknowledgements}
We thank the reviewers in advance for their comments.

\begin{supplement}
\stitle{SUPPLEMENTARY MATERIALS (Valid and Approximately Valid Confidence Intervals for Current Status Data)}
\slink[url]{DOI: .pdf}
\sdescription{The supplement contains mathematical details and figures.
}

\end{supplement}


\clearpage
\setcounter{page}{1}
\setcounter{section}{0}
\setcounter{figure}{0}
\setcounter{equation}{0}

\begin{frontmatter}

\title{SUPPLEMENTARY MATERIALS (Valid and Approximately Valid Confidence Intervals for Current Status Data)}
\runtitle{Valid and Approx. C.I.s for Current Status Data}


\author{\fnms{Sungwook} \snm{Kim}\ead[label=e1]{s.kim@usciences.edu}},
\author{\fnms{Michael P.} \snm{Fay}\ead[label=e2]{mfay@niaid.nih.gov}}
\and
\author{\fnms{Michael A.} \snm{Proschan}
\ead[label=e3]{proscham@niaid.nih.gov}
}

\runauthor{S.Kim, M.P.Fay, and M.A.Proschan}




\end{frontmatter}

\renewcommand{\thesection}{S\arabic{section}}
\renewcommand{\thefigure}{S\arabic{figure}}

\mathversion{bold}
\section{A Specific Form of $a^{*}$ and $b^{*}$}
\mathversion{normal}
\label{Appendix_Specific_forms_a_b}

When the lower confidence limit exceeds the upper confidence limit, we abandon using separate proportions for the lower and upper intervals.  Instead, we use a single proportion for the $m/2$ observations less than, and $m/2$ observations greater than, $t$. To be precise, we do the following.
If $G$ is discrete, then we define $a^{*}$ and $b^{*}$ to be
\begin{equation}
\label{definitionabStar_Discrete}
\begin{array}{l}
a^{*}\equiv a^{*}(t,n,\bold{C})= \left\{
  \begin{array}{l l}
    0& \quad \text{if $C_{\lceil (m-J)/2 \rceil}\geq C_{g}$ }\\
    C_{l-\lceil (m-J)/2 \rceil+1-J} & \quad  \text{if $C_{\lceil (m-J)/2 \rceil}< C_{g}$ and $m>J$}\\
   t & \quad  \text{if $C_{\lceil (m-J)/2 \rceil}< C_{g}$ and $J\geq m$};
  \end{array} \right.\\
$~$\\
b^{*}\equiv b^{*}(t,n,\bold{C})= \left\{
  \begin{array}{l l}
    \infty& \quad \text{if $C_{n-\lceil (m-J)/2 \rceil+1}\leq C_{l}$ }\\
    C_{g+\lceil (m-J)/2 \rceil-1+J} & \quad \text{if $C_{n-\lceil (m-J)/2 \rceil+1}> C_{l}$  and $m>J$}\\
  t & \quad \text{if $C_{n-\lceil (m-J)/2 \rceil+1}> C_{l}$ and $J\geq m$}
  \end{array} \right.
\end{array}
\end{equation}
where $J$ is the number of observations at $t$, and $\lceil (m-J)/2 \rceil$  is the smallest integer greater than or equal to $(m-J)/2$.
If $G$ is continuous, then we define $a^{*}$ and $b^{*}$ to be
\begin{equation}
\label{definitionabStar}
\begin{array}{l}
a^{*}\equiv a^{*}(t,n,\bold{C})= \left\{
  \begin{array}{l l}
    0& \quad \text{if $C_{\lceil m/2 \rceil}>t$ }\\
    C_{l-\lceil m/2 \rceil+1} & \quad \text{if $C_{\lceil m/2 \rceil}\leq t$};
  \end{array} \right.\\
$~$\\
b^{*}\equiv b^{*}(t,n,\bold{C})= \left\{
  \begin{array}{l l}
    \infty& \quad \text{if $C_{n-\lceil m/2 \rceil+1}<t$ }\\
    C_{g+\lceil m/2 \rceil-1} & \quad \text{if $C_{n-\lceil m/2 \rceil+1}\geq t$}
  \end{array} \right.
\end{array}
\end{equation}
where $\lceil m/2 \rceil$  is the smallest integer greater than or equal to $m/2$.

\mathversion{bold}
\section{mid-P Binomial Intervals}
\mathversion{normal}
\label{Appendix_midP}

 In general, the Clopper-Pearson interval for the binomial is conservative, and
 the actual confidence level exceeds the nominal confidence level $(1-\alpha)$ for almost all values of the parameter in order not to be less than $(1-\alpha)$ for any.  To eliminate the conservativeness, one approach is to use the mid-$P$ method
 \citep{lancaster1961significance}. For discrete data, a valid p-value can be calculated as the probability (maximized under the null hypothesis model) of observing equal or more extreme data. The mid-$P$ value slightly adjusts this and is the probability of observing more extreme data plus {\it half} the probability of observing equally extreme data.
The mid-$P$ $100(1-\alpha)\% > 50\%$ confidence limits for a binomial parameter, $\theta$, assuming $Y\sim\text{Binomial}(N,\theta)$ can be found by solving equations for fixed $y$ and $n$:
\[ U_{mid} (1-\alpha/2; y,n) = \left\{ \begin{array}{ll}
       \theta: \text{Pr}(Y<y; \theta)+(.5)\text{Pr}(Y=y;\theta)=\alpha/2 & \mbox{if $y <n$};\\
        1 & \mbox{if $y =n$},\end{array} \right. \]
and
\[ L_{mid}(1-\alpha/2; y,n) = \left\{ \begin{array}{ll}
       \theta: \text{Pr}(Y>y;\theta)+(.5)\text{Pr}(Y=y;\theta)=\alpha/2 & \mbox{if $y>0$};\\
        0 & \mbox{if $y =0$}.\end{array} \right. \]

\section{SMLE of $F$}

\cite{groeneboom2010maximum} defined  the SMLE $\hat{F}_{n}(t)$   for the true $F(t)$ by
\begin{equation}
\label{Fn_with_kernel}
\hat{F}^{SM}_{n}(t)=\int K_{h}(t-u)d\hat{F}_{n}(u);
\end{equation}
 the SMLE $\hat{f}_{n}(t)$   for the true $f(t)$ by
\begin{equation}
\label{f_with_kernel}
\hat{f}^{SM}_{n}(t)=\int k_{h}(t-u)d\hat{F}_{n}(u),
\end{equation}
where $k$ is a triweight kernel which is symmetric and twice continuously differentiable on
$[-1,1]$,
 $K(t)=\int^{t}_{-\infty}k(u)du$,  $K_{h}(u)=K(u/h)$, $k_{h}(u)=(1/h)k(u/h)$, $\hat{F}_{n}(u)$ is the nonparametric maximum likelihood estimator (NPMLE),  and $h>0$ is the bandwidth.

$Theorem~4.2$ in \cite{groeneboom2010maximum} showed that for fixed $t>0$, the asymptotic mean squared error (aMSE)-optimal value of $h$ for estimating $F(t)$ is given by $h_{n,F}=c_{F}n^{-1/5}$, where
\begin{equation}
\label{opti_c}
c_{F}=\left[\frac{F(t)\{1-F(t)\}}{g(t)}\int k(u)^{2}du\right]^{1/5}\times\left[\left\{\int u^{2}k(u)du\right\}^{2}f^{\prime}(t)^{2}\right]^{-1/5},
\end{equation}
 $f^{\prime}(t)$ is the first derivative of $f(t)$.
However the aMSE depends on the unknown distribution $F$, so $c_{F}$ and $h_{F}$ are unknown. Therefore we cannot use (\ref{opti_c}) for estimating $F(t)$ in practice.

To overcome this problem, \cite{groeneboom2010maximum}  introduced the smoothed bootstrap for $F(t)$. They set  the initial choice of the bandwidth, $h_{0}=c_{0}n^{-1/5}$ for $F(t)$, then sampled $m^{\prime}$ observations $(m^{\prime}\leq n)$ from the distribution SMLE $\hat{F}_{n,h_{0}}^{SM}$. They determined the estimator $\hat{F}_{n,cm^{-1/5}}^{SM}$, then repeated $B$ times (they set $B$=500), and estimated aMSE(c) by
 \[
\widehat{MSE}_{B}(c)=B^{-1}\sum^{B}_{i=1}\left(\hat{F}_{n,cm^{-1/5}}^{SM,i}(t)-\hat{F}_{n,h_{0}}^{SM}(t)\right)^{2}.
\]
They defined $\hat{c}_{F,SM}$ as the minimizer of $\widehat{MSE}_{B}(c)$ and then estimated the optimal bandwidth by $\hat{h}_{n,F,SM}= \hat{c}_{F,SM}n^{-1/5}$.

 In this paper, we estimate $f^{\prime}(t)$, then estimate $F(t)$ and $f(t)$ without utilizing bootstrap sampling or Monte Carlo simulation. \cite{groeneboom2010maximum} used the triweight kernel $k(t)=(35/32)(1-t^{2})^{3}1_{[-1,1]}(t)$, but we use the Gaussian kernel for $F(t)$ and $f(t)$.  Other well-known kernels are also applicable to estimate $F(t)$ and $f(t)$. We also estimate $g(t)$ by the kernel density estimation with the Gaussian kernel.

To estimate $g(t)$,
we use the bandwidth recommended by  \cite{silverman1986density}:
\[
 \hat{h}_{n,g}=.9~\text{min}(s, IQR/1.34)n^{-1/5}
\]
where $s$ and $IQR$  are the sample standard deviation  and  sample interquartile range of the  $C_i$ values.
Then the initial $F(t)$, $\hat{F}^{\text{Initial}}(t)$, is estimated by (\ref{Fn_with_kernel}) and $f^{\prime}(t)$ is estimated by
\begin{equation}
\label{f_pp_with_kernel}
\hat{f}^{\prime}_{n}(t)=\int k^{\prime}_{h}(t-u)d\hat{F}_{n}(u),
\end{equation}
with the initial  $h$, say $\hat{h}^{\text{Initial}}_{n,F}$
set to $h_{n,g}$,
where $k^{\prime}_{h}(u)=(1/h^{2})k^{\prime}(u/h)$ and $k^{\prime}(u)$ is the first derivative of $k(u)$.  Then $\hat{c}^{\text{New}}_{n,F}$ and $\hat{h}^{\text{New}}_{n,F}$ are calculated  by substituting $\hat{F}^{\text{Initial}}(t)$, $\hat{f}^{\prime}_{n}(t)$ and $\hat{g}(t)$ into (\ref{opti_c}). Note that if  $\hat{F}^{\text{Initial}}(t)=0$ or 1, $\hat{f}^{\prime}_{n}(t)=0$, or $\hat{g}(t)=0$, then  (\ref{opti_c}) is zero or undefined. Therefore, we need to modify them such as  $(0+\epsilon)\leq\hat{F}^{\text{Initial}}(t)\leq(1-\epsilon)$, $\{\hat{f}^{\prime}_{n}(t)\}^{2}\geq \epsilon$, and $\hat{g}(t)\geq \epsilon$ where $\epsilon$ is a small positive value.  We modify them such that if $\hat{F}^{\text{Initial}}(t)\leq$ .01, set $\hat{F}^{\text{Initial}}(t)=.01$, if $\hat{F}^{\text{Initial}}(t)\geq$ .99, set $\hat{F}^{\text{Initial}}(t)=.99$, if $\{\hat{f}^{\prime}_{n}(t)\}^{2}$$\leq$ $10^{-3}$, set $\{\hat{f}^{\prime}_{n}(t)\}^{2}=10^{-3}$, and if $\hat{g}(t)\leq 10^{-4}$, set $\hat{g}(t)=10^{-4}$.
With $\hat{h}^{\text{New}}_{n,F}$, $F(t)$ is estimated again by (\ref{Fn_with_kernel}), and  $f(t)$ is estimated by (\ref{f_with_kernel}).

We do not iterate this process until convergence, because the iteration of this process does not  guarantee convergence to the true values. We also performed \cite{groeneboom2010maximum}'s smoothed bootstrap with $h_{0}= \hat{h}^{\text{Initial}}_{n,F}$ and compared with our method. The two methods showed very similar estimates, but our method is much faster than smoothed bootstrapping, so we do not present the latter method.

An additional practical adjustment was needed.
Note that if  $\hat{F}^{SM}_{n}(t)=0$ or 1,  $\hat{f}^{SM}_{n}(t)=0$, or $\hat{g}(t)=0$, then $m^{\dagger*}_{n}$ is zero or undefined. Therefore, we  modify them such that
if $\hat{F}^{SM}_{n}(t)$ $\leq$ .01, set $\hat{F}^{SM}_{n}(t)=.01$, if $\hat{F}^{SM}_{n}(t)$ $\geq$ .99, set $\hat{F}^{SM}_{n}(t)=.99$,  if $\hat{f}^{SM}_{n}(t)\leq 10^{-4}$, set $\hat{f}^{SM}_{n}(t)=10^{-4}$, and  if $\hat{g}(t)\leq 10^{-4}$, set $\hat{g}(t)=10^{-4}$.


\section{Figures}

In the following pages are supplemental figures.

\begin{figure}[h]
\centering
\includegraphics[height=5in,width=5in]{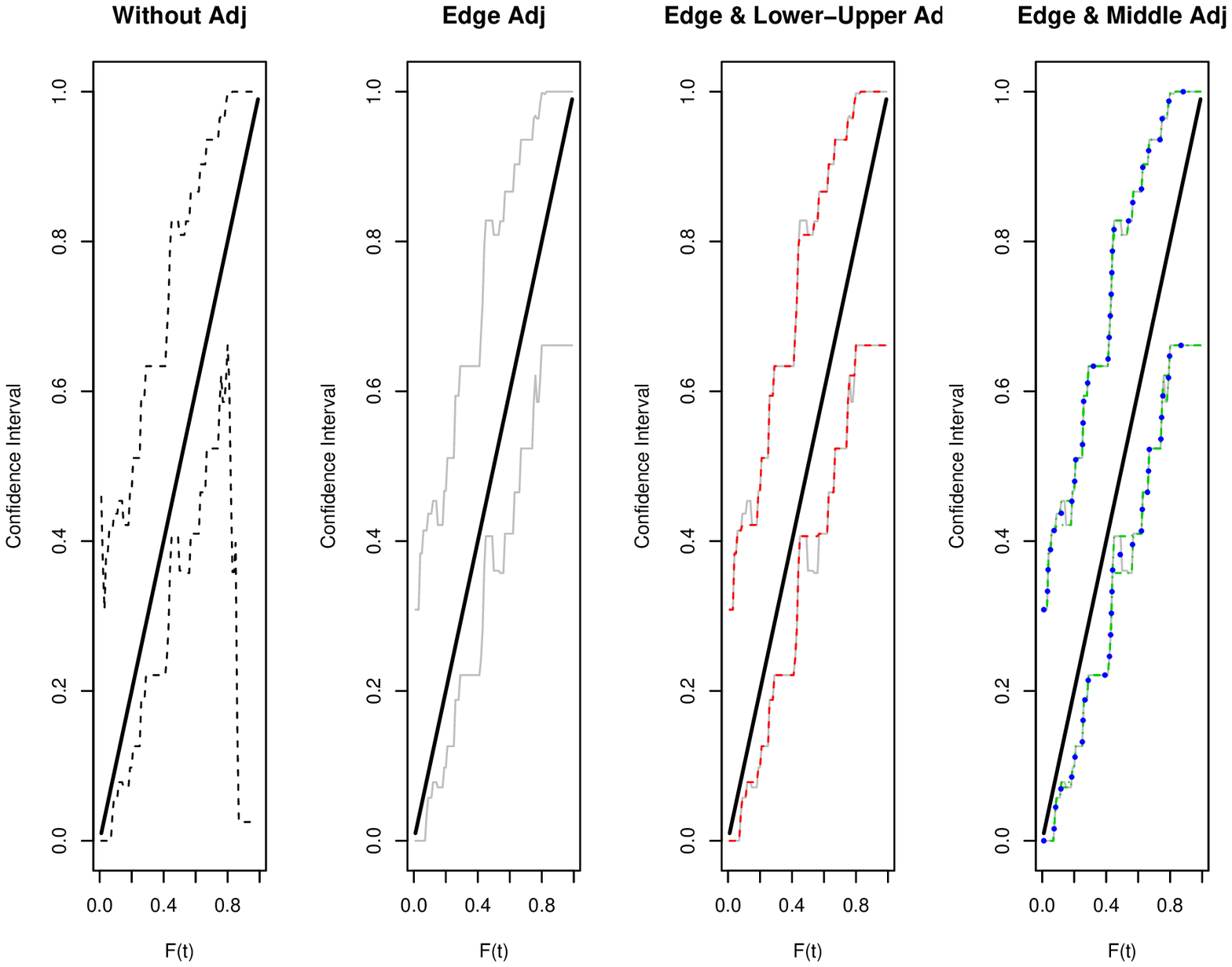}
 \label{FigureAdjustments}
 \caption{An example about the confidence intervals with adjustments. The true $F(t)$ (black solid line), CIs with the edge adjustment (gray solid line), CIs without adjustment (black-dashed line in the first panel), CIs with the edge and lower-upper adjustment (red-dashed line in the third panel), CIs with the edge and middle value adjustment (blue-dotted line in the fourth panel). $f(t)$ is  Gamma(3,1),  $g(t)$ is  Unif(0,5), and $n$=50.}
\end{figure}

\clearpage
\begin{figure}[h]
\centering
\includegraphics[height=4in,width=5in]{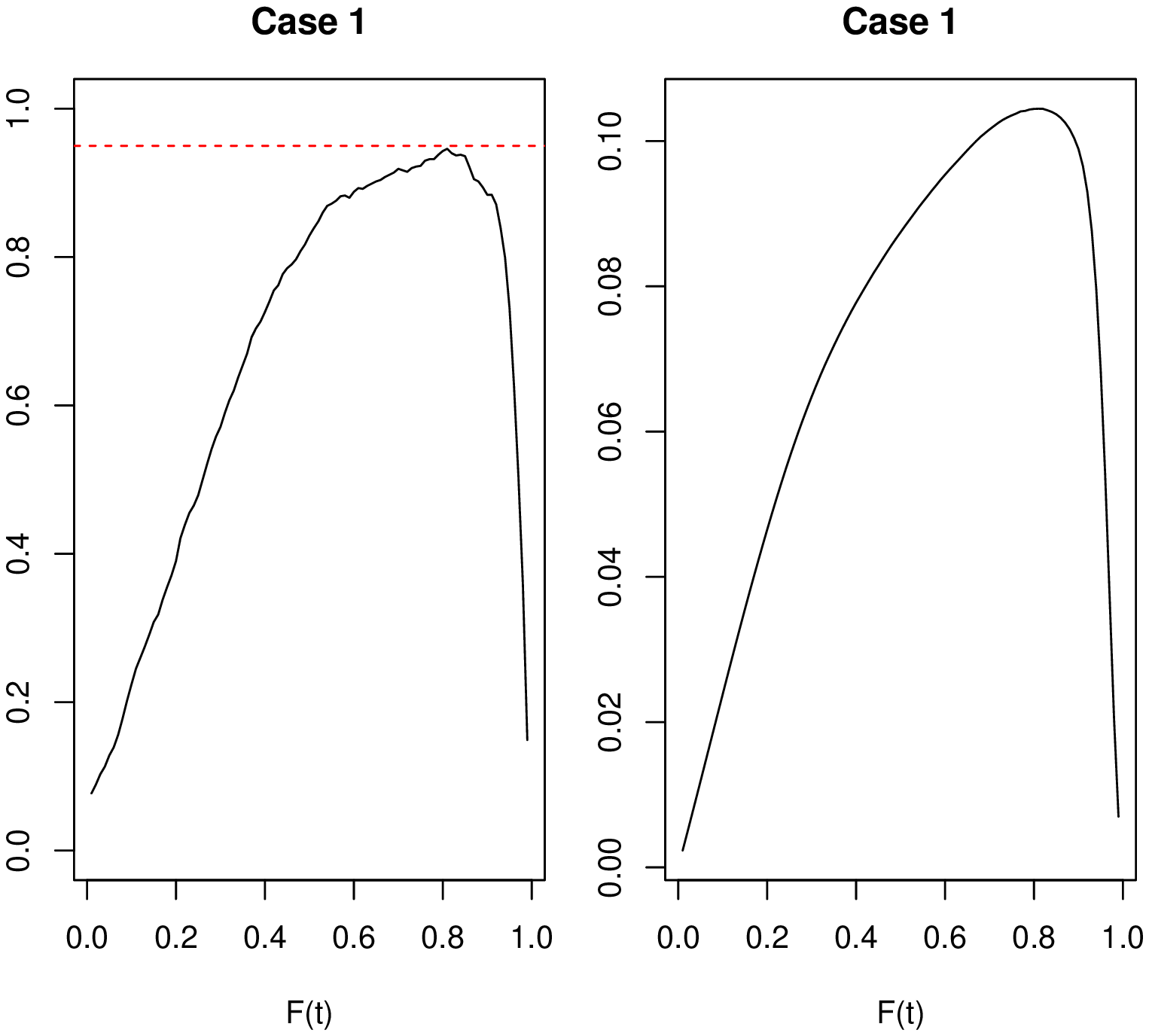}
  \label{SMLE_CI_1000}
\caption{Simulated coverage and  average length of  95 \% SMLE-based confidence intervals for Case 1.    We set $a=0$ and $b$ is the maximum of the sample (assessment times) in the equation (9.75)
\citep[see][]{groeneboom2014nonparametric}.
We used the triweight kernel, and set the bandwidth, $h=(F^{-1}(.99))n^{-1/4}$. The sample size is $n$=1,000. For each sample, we generated 1,000 bootstrap samples, and computed the 20th and 980th percentile of the values (9.76). Then we computed (9.77) in
\cite{groeneboom2014nonparametric}.
1,000 replications have been performed.}
\end{figure}

\clearpage
\begin{figure}[h]
\centering
\includegraphics[height=4in,width=5in]{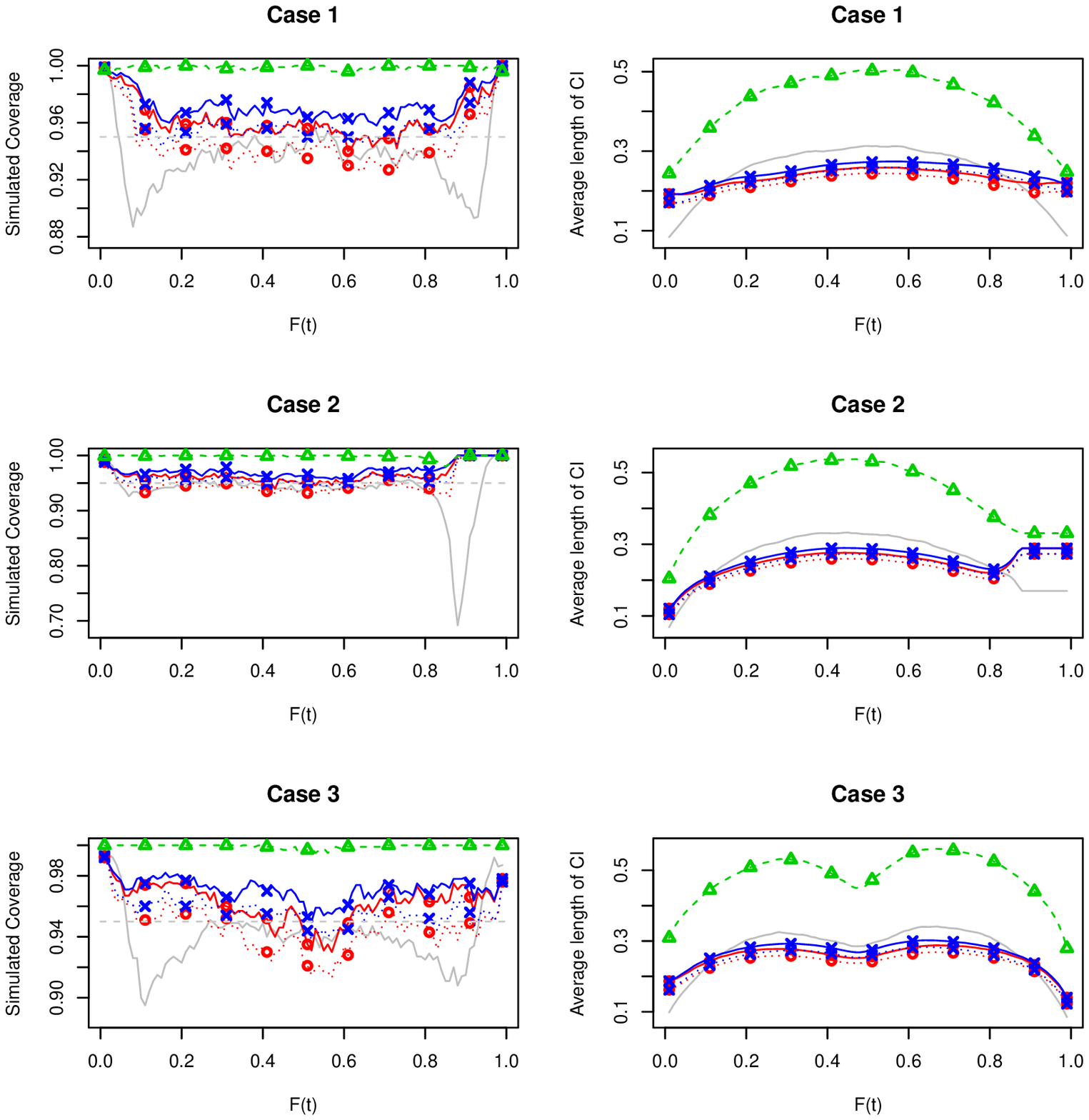}
  \label{Figure_Six_Plots_200}
\caption{Simulated coverage and average length of 95 \% confidence intervals for Case 1,2, and 3. The likelihood ratio-based CI (gray solid line);  the ABF CI with edge and lower-upper adjustment (red solid line with ($\circ$)); the mid-$P$ ABF CI with edge and lower-upper adjustment (red dotted line with ($\circ$));   the ABF CI with edge and middle value adjustment (blue solid line with ($\times$)); the mid-$P$ ABF CI with edge and middle value adjustment (blue dotted line with ($\times$));   the valid CI  with $m_{n}=n^{2/3}$ (green dashed line with ($\triangle$)).  The sample size is $n$=200, and 1,000 replications have been performed.}
\end{figure}

\clearpage
\begin{figure}[h]
\centering
\includegraphics[height=4in,width=5in]{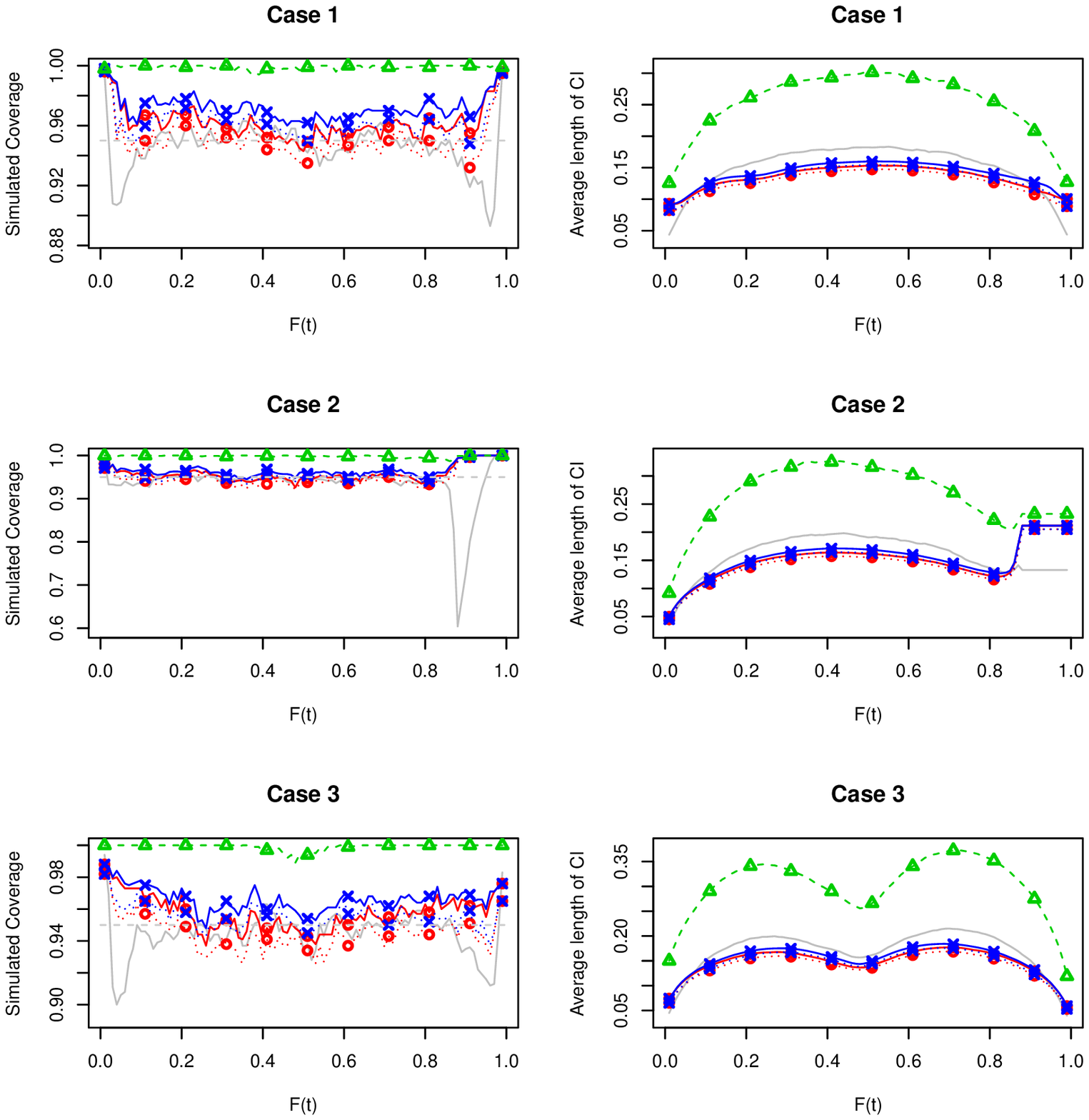}
  \label{Figure_Six_Plots_1000}
\caption{Simulated coverage and  average length of 95 \% confidence intervals for Case 1,2, and 3. The likelihood ratio-based CI (gray solid line);  the ABF CI with edge and lower-upper adjustment (red solid line with ($\circ$)); the mid-$P$ ABF CI with edge and lower-upper adjustment (red dotted line with ($\circ$));   the ABF CI with edge and middle value adjustment (blue solid line with ($\times$)); the mid-$P$ ABF CI with edge and middle value adjustment (blue dotted line with ($\times$));   the valid CI  with $m_{n}=n^{2/3}$ (green dashed line with ($\triangle$)).  The sample size is $n$=1,000, and 1,000 replications have been performed.}
\end{figure}

\clearpage
\begin{figure}[h]
        \centering
        \includegraphics[height=5in,width=5in]{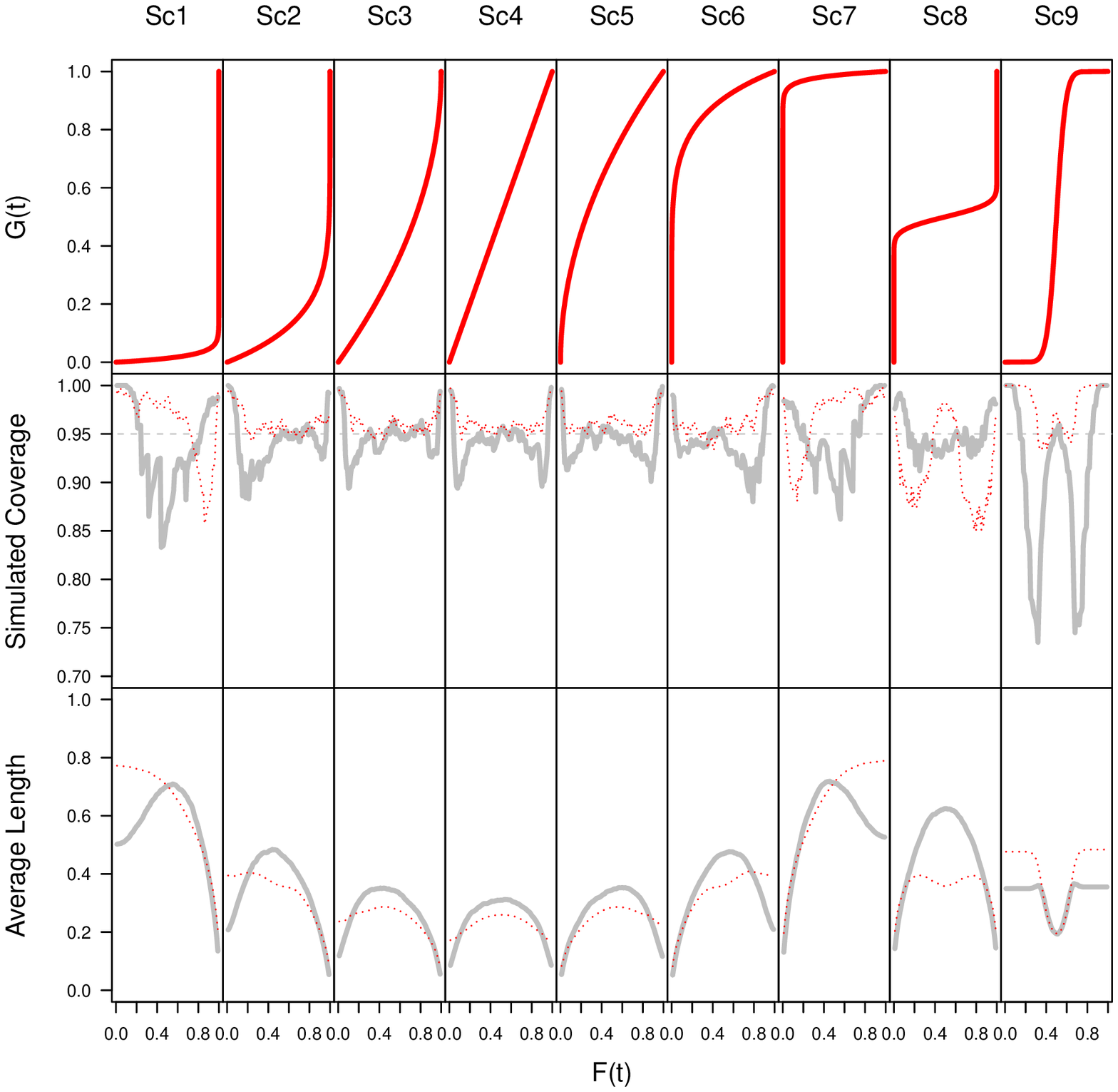}
\caption{Simulated coverage and averaged length of 95 \% CIs  : the likelihood ratio-based CI (gray solid line); the mid-$P$ ABF CI with edge and middle value adjustment (red dotted line) with $n$ = 200. There are 9 scenarios which are described in text, and simulations
based on 1,000 replications.}
        \end{figure}

\clearpage
\begin{figure}[h]
        \centering
        \includegraphics[height=5in,width=5in]{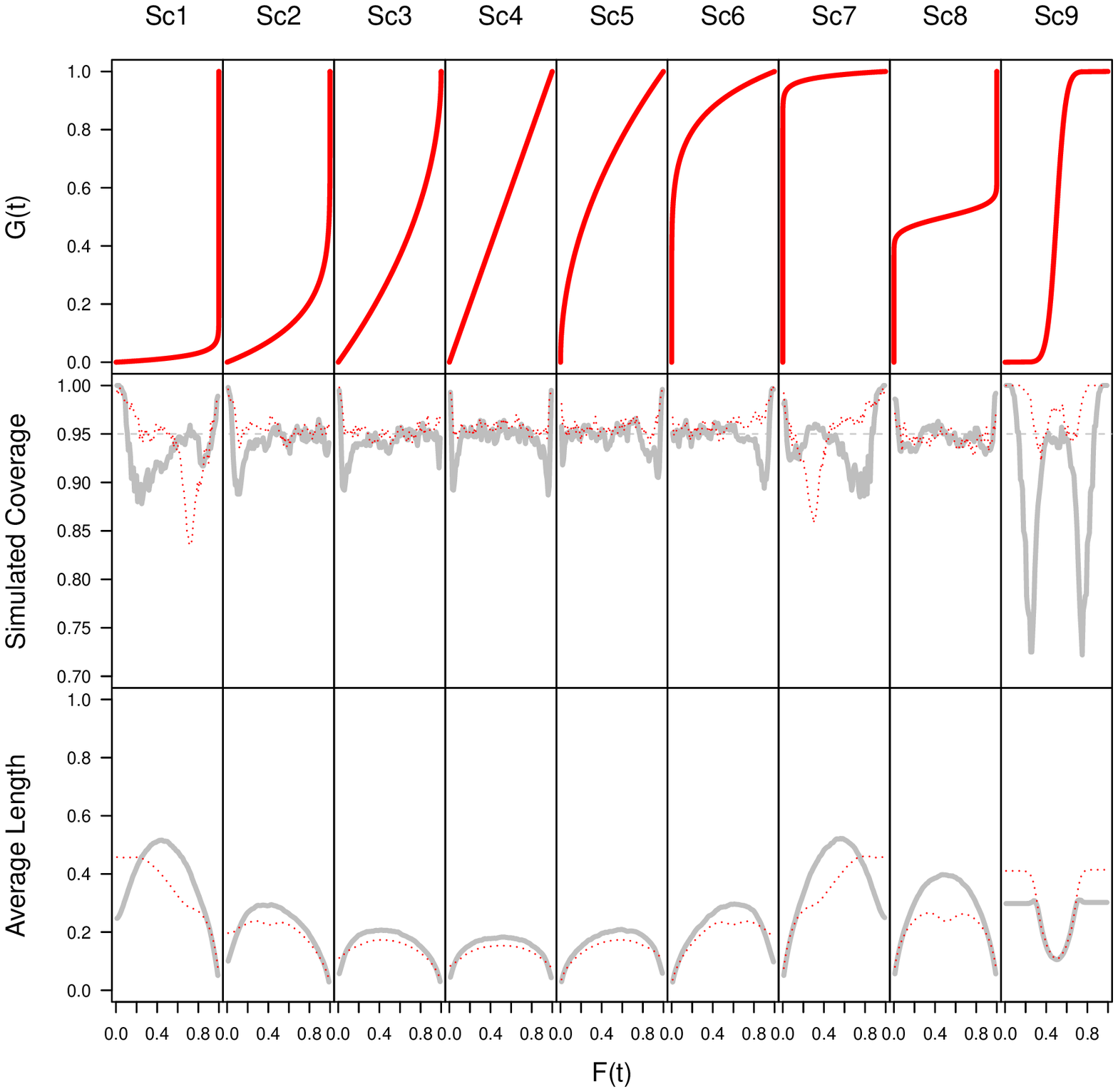}
\caption{Simulated coverage and averaged length of 95 \% CIs  : the likelihood ratio-based CI (gray solid line); the mid-$P$ ABF CI with edge and middle value adjustment (red dotted line) with $n$ = 1000. There are 9 scenarios which are described in text, and simulations
based on 1,000 replications.}
  \end{figure}

\clearpage
\begin{figure}[h]
        \centering
       \includegraphics[height=5in,width=5in]{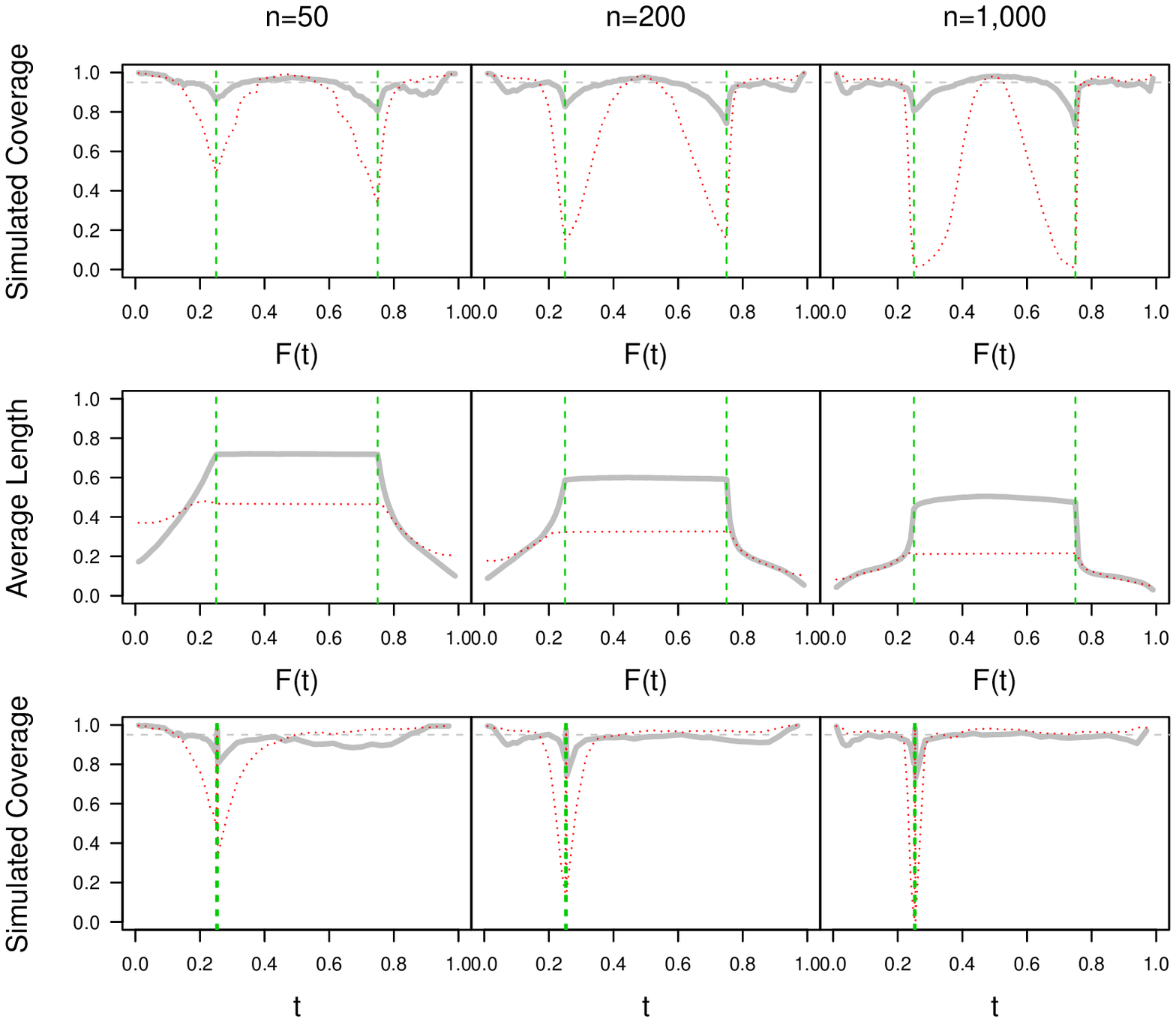}
\caption{Simulated coverage and averaged length of 95 \% CIs  : the likelihood ratio-based CI (gray solid line); the mid-$P$ ABF CI with edge and middle value adjustment (red dotted line). Simulations
are based on 1,000 replications.
 $g\sim \text{Unif}(0,1)$;}
 \begin{equation*}
F(t)= \left\{
  \begin{array}{l l}
    t& \quad \text{for $t\leq .25$ };\\
    .25+(20,000)(t-.25)^2 & \quad \text{for $.25<t\leq \left(.25+\frac{1}{200}\right)$};\\
    .75+\frac{.25}{(.75-\frac{1}{200})}(t-.25-\frac{1}{200}) & \quad \text{for $\left(.25+\frac{1}{200}\right)<t\leq 1$}.
  \end{array} \right.
  \end{equation*}
\end{figure}

\end{document}